\documentclass[12pt]{article}
\usepackage{graphicx}
\usepackage{amssymb}
\usepackage{amsmath}
\usepackage{amscd}
\usepackage{color}
\usepackage{url}

\def\code#1{{\tt #1}}

\title{Mesh Algorithms for PDE with Sieve I: Mesh Distribution}
\author{Matthew G. Knepley\\Computation Institute\\University of Chicago\\Chicago, IL 60637 \and
        Dmitry A. Karpeev\\Mathematics and Computer Science Division\\Argonne National Laboratory\\Argonne, IL 60439}

\begin{document}
\maketitle

\begin{abstract}
We have developed a new programming framework, called Sieve, to support parallel numerical PDE\footnote{Partial differential equation(s).}
algorithms operating over distributed meshes.  We have also developed  a reference implementation of Sieve in C++
as a library of generic algorithms operating on distributed containers conforming to the Sieve interface.
Sieve makes instances of the incidence relation, or \emph{arrows}, the conceptual first-class objects
represented in the containers.  Further, generic algorithms acting on this arrow
container are systematically used to provide natural geometric operations on the topology and also, through duality, on the data.
Finally, coverings and duality are used to encode not only individual meshes, but all types of hierarchies underlying 
PDE data structures, including multigrid and mesh partitions. 

In order to demonstrate the usefulness of the framework, we show how the mesh partition data can be represented
and manipulated using the same fundamental mechanisms used to represent meshes. We present the
complete description of an algorithm to encode a mesh partition and then distribute a mesh, which is independent 
of the mesh dimension, element shape, or embedding. 
Moreover, data associated with the mesh can be similarly distributed with exactly the same
algorithm. The use of a high level of abstraction within the Sieve leads to several benefits in terms of code reuse,
simplicity, and extensibility.  We discuss these benefits and compare our approach to other existing mesh libraries.
\end{abstract}

\section{Introduction}

Numerical PDE codes frequently comprise of two uneasily coexisting pieces: the mesh, describing the topology
and geometry of the domain, and the functional \emph{data} attached to the mesh representing the discretized 
fields and equations.  The mesh data structure typically reflects the representation used by the mesh 
generator and carries the embedded geometic information.  
While this arrangement is natural from the point of view of mesh generation and exists in the best of such packages
(e.g., \cite{shewchuk96}), it is frequently foreign to the process of solving equations on the generated mesh. 

At the same time, the functional data closely reflect the linear algebraic structure of the computational 
kernels ultimately used to solve the equations; here the natural geometric structure of the equations, 
which reflects the mesh connectivity in the coupling between the degrees of freedom, is sacrificed to the 
rigid constraints of the solver.  In particular, the most natural geometric operation of a restriction 
of a field to a local neighborhood entails tedious and error-prone index manipulation.  

In response to this state of affairs a number of efforts arose addressing the fundamental issues of interaction 
between the topology, the functional data and algorithms.  We note the MOAB project \cite{TautgesMeyers04,Tautges04,Meyers02}
and the TSTT/ITAPS SciDAC projects \cite{Meyers02,SeolShepard05,BeallWalshShephard04}, 
the libMesh project \cite{CareyAndersonCarnesKirk04}, the GrAL project \cite{GrALThesis}, to name just a few.  Sieve
shares many features with these projects, but GrAL is the closest to it in spirit. Although each of these projects
addresses some of the issues outlined above, we feel that there is room for another approach.

Our Sieve framework, is a collection of interfaces and algorithms for manipulating geometric data.  The design
may be summarized by considering three constructions.
First, data in Sieve are indexed by the underlying geometric elements, such as mesh cells, rather than by 
some artificial global order.  Further, the local traversal of the data is based 
on the connectivity of the geometric elements.  For example, Sieve provides operations that, given a mesh cell, 
traverse all the data on its interior, its boundary,  or its closure. Typical
operations on a Sieve are shown in Table~\ref{table:SieveOps} and described in greater detail in
Section~\ref{subsec:containers}. In the table, topological mesh elements, such as vertices, edges, and so on, are refered to
as abstract \emph{points}
\footnote{Our \emph{points} correspond to geometric \emph{entities} in some other approaches like MOAB or ITAPS}, 
and the adjacency relation between two points, such as an edge and its vertex, is refered to as
\emph{covering}: an edge is coverted by its end vertices. Notice that exactly the same operation is used to obtain edges
adjacent to a face as faces adjacent to a cell, without even a lurking dimension parameter. This is the \emph{key} to
enabling dimension-independent programming for PDE algorithms.

Second, the global topology is divided into a chain of local topologies with an overlap structure relating them to each
other.  The overlap is encoded using the Sieve data structure again, this time containing arrows relating points in
different local topologies. 
The data values over each local piece are manipulated using the local connectivity, and each local piece may
associate different data to the same global element. The crucial ingredient here is the operation of
assembling the chain of local data collections into a consistent whole over the global topology.

Third, the covering arrows can carry additional information, controlling the way in which the data from the covering points
are assembled onto the covered points. For example, orientation information can be encoded on the arrows to dictate an
order for data returned over an element closure. More sophisticated operations are also possible, such as linear
combinations which enable coordinate transformations, or the projection and interpolation necessary for multigrid algorithms.
This is the central motivation behind the arrow-centric interface.

Emphasis on the covering idea stems directly from the cell complex construction in algebraic topology. We have
abstracted it along the lines of category theory, with its emphasis on arrows, or morphisms, as the organizing
principle. The analogy runs deeper, however, because in PDE applications meshes do not exist for their own sake, but to
support geometrically structured information.  The geometric structure of these data manifests itself through duality
between topogical operations, such as \emph{closure} of a mesh element, and analytical operations, such as the
\emph{restriction} of a field to a closed neighborhood of the element.  Formally this can be seen as a reversal of
arrows in a suitable category.  At the practical level, this motivates the arrow-centric point of view, which
allows us to load the arrows with the data (e.g., coordinate transformation parameters) making the dualization between
covering and restriction possible.

The arrow-centric point of view also distinguishes our approach from similar projects such as \cite{GrALThesis}.
In addition, it is different from the concept of a flexible database of geometric entities underlying the MOAB and
TSTT/ITAPS methodologies (see e.g., \cite{TautgesMeyers04} and \cite{SeolShepard05}).  Sieve can be thought of as a database, 
but one that limits the flexibility by insisting on the arrow-centric structure of the input and output and a small 
basic query interface optimized to reveal the covers of individual elements. 
This provides a compact conceptual universe shifting the flexibility to the generic algorithms enabled by a
well-circumscribed container interface.  

Although other compact interfaces based on a similar notion \emph{adjacency} exist, we feel that Sieve's interface and
the notion of a covering better capture the essense of the geometric notions underlying meshes, rather than mapping them
onto a database-inspired language.  Moreover, these adjacency queries often carry outside information, such as dimension
or shape tags, which is superfluous in the Sieve interface and limits the opportunity for dimension independent
programming. These geometric notions are so universal that the systematic use of covering notions is possible at all
levels of hierarchy underlying PDE computation.  For example, the notion of covering is used to record relations between
vertices, edges and cells of other dimensions in a sieve.  No separate relation is used to encode ``side'' adjacencies,
such as ``neighbor'' relations between cells of the same dimension, as is done in GrAL.  

In fact, the points of a sieve are not a priori interpreted as elements of different dimensions
and covering can be used to encode \emph{overlap} relations in multiple non-conforming meshes, multigrid hierarchies,
or even identification of cells residing on multiple processors.  Contrast this, for example, with the multiple
notions employed by ITAPS to describe meshes: meshes, submeshes, mesh entities, mesh entity sets and parallel mesh
decompositions.  While the relations between all these concepts are of essentially similar nature, this unity
is not apparent in the interface, inhibiting reuse and hindering analysis of the data structures, their capabilities
and their complexity.

Undoubtedly, other approaches may be more appropriate in other computational domains.  For instance, different
data structures may be more appropriate for mesh generation, where very different types of queries, modifications
and data need to be associated with the mesh.  Partitioning algorithms may also require different data access patterns
to ensure efficiency and scalability.  Sieve does not pretend to address those concerns.  Instead, we try to focus
on the demands of numerical PDE algorithms that revolve around the idea of a field defined over a geometry.
Different PDE problems use different fields and even different numbers of fields with different discretizations.
The need for substantial flexibility in dealing with a broad class of PDE problems and their geometric nature 
are the main criterion for the admission into the Sieve interface.

Here we focus on the reuse of the basic covering notions at different levels of data hierarchy.  In
particular, the division of the topology into pieces and assembly over an overlap is among the fundamental notions of
PDE analysis, numerical or otherwise.  It is the essence of the domain decomposition method and can be used in parallel
or serial settings, or both.  Moreover, we focus on this decomposition/assembly aspect of Sieve and present its
capabilities with a fundamental example of this kind --- the distribution of a mesh onto a collection of processors.  It
is a ubiquitous operation in parallel PDE simulation and a necessary first step in constructing the full distributed
problem.  Moreover mesh distribution makes for an excellent pedagogical problem, illustrating the powerful simplicity of
the Sieve construction. The Sieve interface allows PDE algorithms, operating over data distributed over a mesh, to be
phrased without reference to the dimension, layout, element shape, or embedding of the mesh. We illustrate this with the
example of distribution of a mesh and associated data fields over it. The same simple algorithm will be used to
distribute an arbitrary mesh, as well as fields of arbitrary data layout.

We discuss not only the existing code for the Sieve library but also the concepts that underlie
its design and implementation. These two may not be in complete agreement, as the code continues to
evolve. We use the \code{keyboard} font to indicate both existing library interfaces and proposed developments
that more closely relate to our design concepts.
Furthermore, early implementations may not be optimal from the point of view of runtime and storage complexity
as we resist premature optimizations in favor of refining the interface.  Nonetheless, our reference implementation
is fully functional, operating in parallel, and in use by real applications~\cite{AagaardKnepleyWilliams05,PCICE}.  This
implementation verifies the viability and the consistency of the interface, but does not preclude more efficient
implementations  better suited to particular uses.  The added value of the interface comes in the enabling of generic
algorithms, which operate on the interface and are independent of the underlying implementation.  In this publication we
illustrate some of these fundamental algorithms.

The rest of the paper is organized as follows.  In Section~\ref{sec:sieve} we introduce the basic notions and algorithms
of the Sieve framework, which are then seen in action in Section~\ref{sec:partition} where the algorithms for mesh distribution and
redistribution in a parallel setting are discussed.  Section~\ref{subsec:distributionEx} contains specific examples of mesh distribution and 
Section~\ref{sec:conclusions} concludes the paper.

\section{Sieve Framework}\label{sec:sieve}
        Sieve can be viewed as a library of parallel containers and algorithms
that extends the standard container collection (e.g., the Standard Template
Library of C++ and BOOST libraries).  The extensions are simple but provide the crucial functionality
and introduce what is, in our view, a very useful semantics.  Throughout this paper
we freely use the modern terminology of generic programming, in particular the idea
of a {\em concept}, which is an interface that a class must implement to be usable 
by templated algorithms or methods.

Our fundamental concept is that of a \code{Map}, which we understand
in the multivalued sense as an assignment of a \code{sequence} of {\em points} in the range
to each of the points in the domain.  A \code{sequence} is an immutable ordered 
collection of points that can be traversed from the \code{begin} element
to the \code{end}. Typically a \code{sequence} has no repetitions, and we assume such
\emph{set} semantics of sequences unless explicitly noted otherwise.  

A \code{sequence} is a basic input and output type 
of most Sieve operations, and the basic operation acting on sequences is called 
\code{restrict}.  In particular, a \code{Map} can be \code{restrict}ed to a
point or a \code{sequence} in the domain, producing the corresponding 
\code{sequence} in the range.  \code{Map} objects can be updated in various ways.
At the minimum we require that a \code{Map} implement a \code{set} operation 
that assigns a \code{sequence} to a given domain point.  Subsequent \code{restrict}
calls may return a \code{sequence} reordered in an implementation-dependent way.

\subsection{Basic containers} \label{subsec:containers}\label{sec:mesh-example}
\code{Sieve} extends the basic \code{Map} concept in several ways.  First, it allows bidirectional
mappings.  Hence we can map points in the range, called the \code{cap}, to
the points in the domain, called the \code{base}.  This mapping is called the \code{support},
while the \code{base}-to-\code{cap} mapping is called the \code{cone}.  

Second, the resulting 
\code{sequence} actually contains not the image points but \code{arrows}.  An \code{arrow} responds 
to \code{source} and \code{target} calls, returning respectively the \code{cap} and \code{base} 
points of the \code{arrow}.  Thus, an \code{arrow} not only abstracts the notion of a pair of points related by the map
but also allows the attachment of nearly arbitrary ``payload'', a capability useful for local traversals.

\begin{figure}
  \begin{center}
  \includegraphics[width=4.5in]{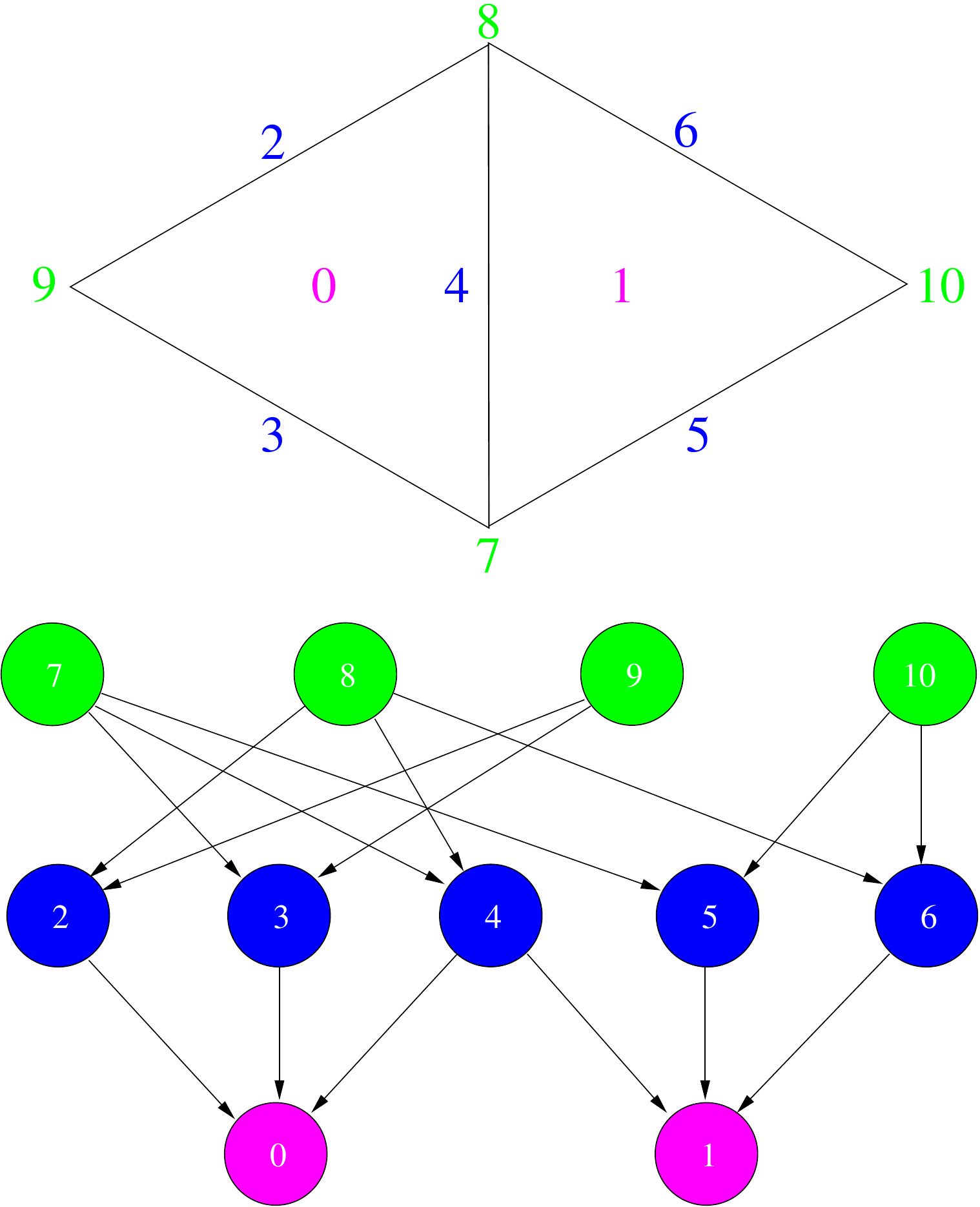}
  \end{center}
  \caption{A simple mesh and its \code{Sieve} representation.}
  \label{fig:doubletSieve}
\end{figure}

One can picture a \code{Sieve} as a bipartite graph with the \code{cap} above the \code{base}
and the \code{arrows} pointing downward (e.g., Fig.~\ref{fig:doubletSieve}).  The containers are not constrained by the type of 
point and \code{arrow} objects, so Sieve must be understood as a library of {\em meta-objects} and {\em meta-algorithms}
(a template library in the C++ notation), which generates appropriate code upon instantiation 
of basis objects.  We primarily have the C++ setting in mind, although appropriate Python and C bindings 
have been provided in our reference implementation.

A \code{Sieve} can be made into a \code{Map} in two different ways, by identifying either \code{cone} or 
\code{support} with \code{restrict}.  Each can be done with a simple adapter class and allows all the basic
\code{Map} algorithms to be applied to \code{Sieve} objects.

The \code{Sieve} also extends \code{Map} with capabilities of more geometric character.
It allows the taking of a transitive closure of \code{cone} to obtain the topological \code{closure} of a point
familiar from cell complex theory~\cite{Hatcher02,Aleksandrov98}.  Here arrows are interpreted as the incidence relations
between points, which represent the cells. Likewise, iterated \code{supports} result in the \code{star}
of a point. The \code{meet(p,q)} lattice operation returns the smallest \code{sequence} of points whose removal would render
\code{closure(p)} and \code{closure(q)} disjoint. The \code{join(p,q)} operation is the analogue for \code{star(p)} and
\code{star(q)}. Note that all these operations actually return \code{arrow} sequences, but by default we extract either
the source or the target, a strategy that aids in the definition of transitive closures and simplifies programming.

\begin{table}
\begin{center}
\begin{tabular}{|l|l|}
\hline
cone(p)    & sequence of points covering a given point p \\\hline
closure(p) & transitive closure of cone \\\hline
support(p) & sequence of points covered by a given point p\\\hline
star(p)    & transitive closure of support \\\hline
meet(p,q)  & minimal separator of closure(p) and closure(q)\\ \hline
join(p,q)  & minimal separator of star(p) and star(q)\\ \hline
\end{tabular}
\caption{Typical operations on a Sieve.}
\label{table:SieveOps}
\end{center}
\end{table}

Fig.~\ref{fig:doubletSieve} illustrates how mesh topology can be represented as a \code{Sieve} object.
The arrows indicate covering or incidence relations between triangles, edges, and vertices of a 
simple simplicial mesh.  Sieve operations allow one to navigate through the mesh topology 
and carry out the traversals needed to use the mesh. We illustrate some common Sieve operations on the mesh
from Fig.~\ref{fig:doubletSieve} in Table~\ref{table:SieveEx}.

\begin{table}
\begin{center}
\begin{tabular}{|l|c|}
\hline
cone(0)    & \{2, 3, 4\} \\\hline
support(4) & \{0, 1\} \\\hline
closure(1) & \{4, 5, 6, 7, 10, 8\} \\\hline
star(8)    & \{2, 4, 6, 0, 1\} \\\hline
meet(0,1)  & \{4\} \\\hline
join(2,4)  & \{0\} \\\hline
join(2,5)  & \{\} \\\hline
\end{tabular}
\caption{Results of typical operations on the Sieve from Fig.~\ref{fig:doubletSieve}.}
\label{table:SieveEx}
\end{center}
\end{table}

\subsection{Data Definition and Assembly}\label{sec:data-assembly}

  \code{Sieves} are designed to represent relations between geometric entities, represented by points. 
They can also be used to attach data directly to \code{arrows}, but not to points, since points may be duplicated 
in different \code{arrows}. A \code{Map}, however,  can be used effectively to lay out data over points.
It defines a \code{sequence}-valued function over the implicitly defined domain \code{set}.  In this case
the domain carries no geometric structure, and most data algorithms rely on this minimal \code{Map} concept.

  \subsubsection{\code{Sections}}\label{subsec:section}
If a \code{Map} is combined with a \code{Sieve}, it allows more sophisticated data traversals such as 
\code{restrictClosure} or \code{restrictStar}.  These algorithms are essentially the composition of maps from points to
point sets (closure) with maps from points to data (section).  Analogous traversals based on \code{meet}, \code{join},
or other geometric information encoded in \code{Sieve} can be implemented in a straightforward manner.
The concept resulting from this combination is called a \code{Section}, 
by analogy with the geometrical notion of a section of a fiber bundle.  Here the \code{Sieve} plays the role
of the base space, organizing the points over which the mapping representing the section is defined.  
We have found \code{Sections} most useful in implementating finite element discretizations of PDE problems.  
These applications of \code{Section} functionality are detailed in an upcoming publication~\cite{KnepleyKarpeev07C}.

  A particular implementation of \code{Map} and \code{Section} concepts ensures contiguous
storage for the values.  We mention it because of its importance for high-performance parallel computing with
Sieve.  In this implementation a \code{Map} class uses another \code{Map} internally that maps
domain points to offsets into a contiguous storage array.  This allows Sieve to interface with parallel
linear and nonlinear solver packages by identifying \code{Map} with the vector from that package. We have done this for
the PETSc~\cite{petsc-user-ref} package. The internal \code{Map} is sometimes called the \emph{atlas} of that \code{Section}.
The analogous geometric object is the local trivialization of a fiber bundle that organizes the 
space of values over a domain neighborhood (see, e.g., \cite{Steenrod99}).

We observe that \code{Sections} and \code{Sieves} are in duality. This duality is expressed by the relation of the
\code{restrict} operation on a \code{Section} to the \code{cone} operation in a \code{Sieve}.
Corresponding to \code{closure} is the traversal of the \code{Section} data implemented by \code{restrictClosure}.
In this way, to any \code{Sieve} traversal, there corresponds a traversal of the corresponding \code{Section}. Pictured
another way, the covering arrows in a \code{Sieve} may be reversed to indicate restriction. This duality will arise
again when we picture the dual of a given mesh in Section~\ref{subsec:dual-graph}.

  \subsubsection{\code{Overlap} and \code{Delta}}\label{subsec:overlap-delta}

In order to ensure efficient local manipulation of the data within a \code{Map} or a \code{Section},
the global geometry is divided into manageable pieces, over which the \code{Maps} are defined.
In the context of PDE problems, the chain of subdomains typically represents local meshes that cover 
the whole domain.  The dual chain, or a cochain, of \code{Maps} represents appropriate restrictions of 
the data to each subdomain.  For PDEs, the cochain comprises local fields defined over submeshes.

The covering of the domain by subdomains is encoded by an \code{Overlap} object.  It can be implemented 
by a \code{Sieve}, whose \code{arrows} connect the points in different subdomains that cover each other.  
Strictly speaking, \code{Overlap} arrows relate  pairs (domain, domain\_point).
Alternatively, we can view \code{Overlap} itself as a chain of \code{Sieves} indexed by nonempty 
overlaps of the subdomains in the original chain.  This better reflects the locality of likely 
\code{Overlap} traversal patterns: for a given chain domain, all points and their covers from other
subdomains are examined.

\begin{figure}
  \begin{center}
  \includegraphics[width=4.5in]{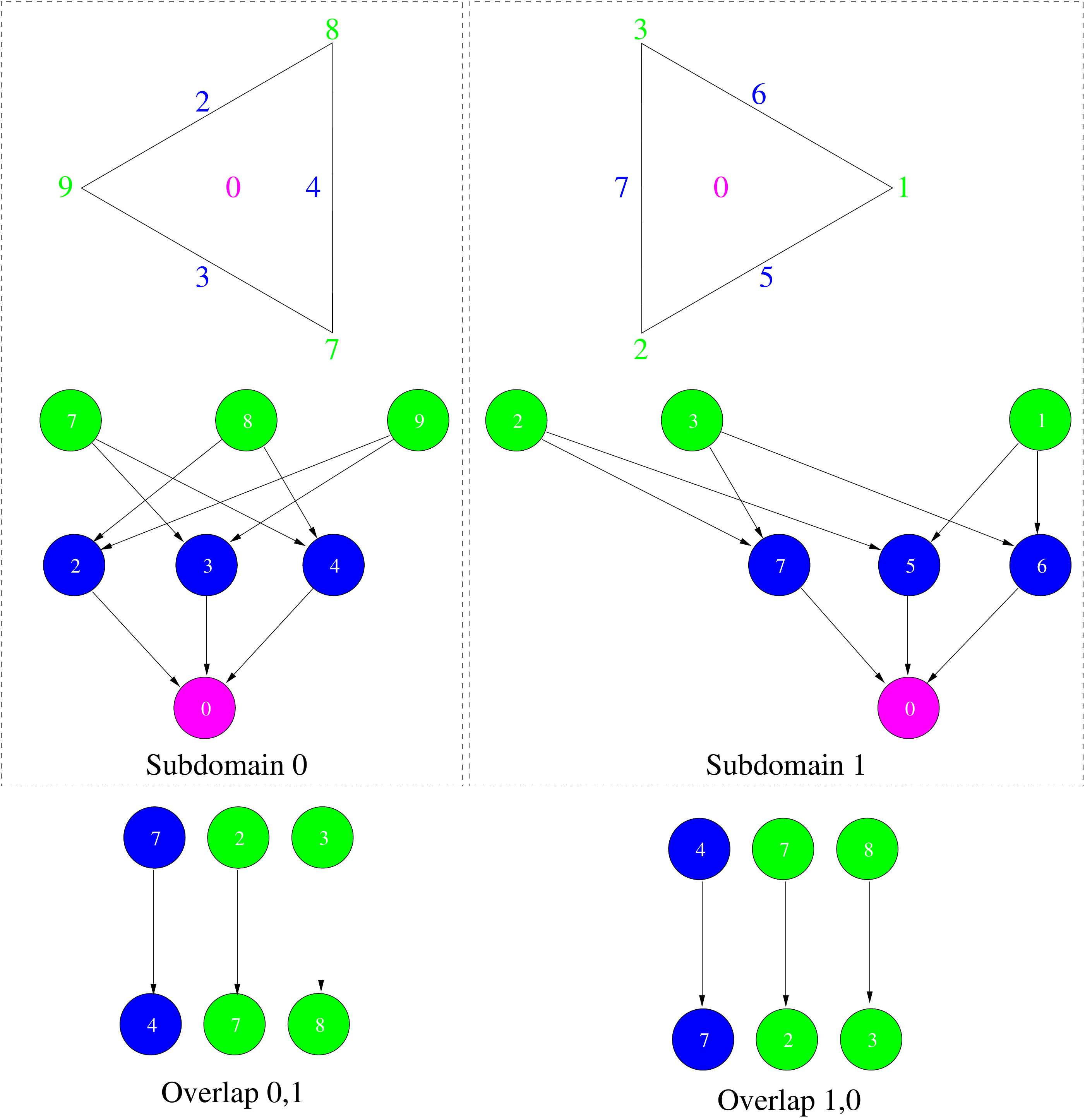}
  \end{center}
  \caption{\code{Overlap} of a conforming mesh chain obtained from breaking up the mesh in Fig.~\ref{fig:doubletSieve}.}
  \label{fig:doubletOverlap}
\end{figure}

An \code{Overlap} is a many-to-many relation.  In the case of meshes this allows for nonconforming
overlapping submeshes. However, the essential uses of \code{Overlap} are evident even in the simplest 
case representing conforming subdomain meshes treated in detail in the example below.
Fig.~\ref{fig:doubletOverlap} illustrates the \code{Overlap} corresponding to a conforming mesh chain
resulting from partitioning of the mesh in Fig.~\ref{fig:doubletSieve}.
Here the \code{Overlap} is viewed as a chain of \code{Sieves}, and the local mesh point
indices differ from the corresponding global indices in Fig.~\ref{fig:doubletSieve}.
This configuration emphasizes the fact that no global numbering scheme is imposed across a chain and the global connectivity
is always encoded in the \code{Overlap}.  In the present case, this is simply a one-to-one identification
relation. Moreover, many overlap representations are possible; the one presented above, while straightforward, differs from
that shown in Section~\ref{sec:serial}.

The values in different \code{Maps} of a cochain are related as well.  The relation among them reflects the overlap 
relation among the points in the underlying subdomain chain.  
The nature of the relationship between values varies according to the problem.  
For example, for conforming meshes (e.g., Fig.~\ref{fig:doubletOverlap}) 
the \code{Overlap} is a one-to-one relation between identified elements of different subdomain meshes.
In this case, the \code{Map} values over the same mesh element in different domains can be duplicates, 
as in finite differences, or partial values that have to be added to obtain the unique global value, as in 
finite element methods. In either case the number of values over a shared mesh element must be the same in the cooverlapping 
\code{Maps}. Sometimes this number is referred to as the {\em fiber dimension}, by analogy with fiber bundles.

Vertex coordinates are an example of a cochain whose values are simply duplicated in different local maps, as shown in
Section~\ref{sec:serial}.
In the case of nonconforming subdomain meshes, \code{Overlap} is a many-to-many relation, and \code{Map} values 
over overlapping points can be related by a nontrivial transformation or a relation.  They can also be different
in number.  All of this information --- fiber dimensions over overlapping points, the details of the data
transformations, and other necessary information --- is encoded in a \code{Delta} class.

A \code{Delta} object can be viewed as a cochain of maps over an \code{Overlap} chain, and is dual to the
\code{Overlap} in the same way that a \code{Section} is dual to a \code{Sieve}.
More important, a \code{Delta} acts on the \code{Map} cochain with domains related by the \code{Overlap}.
Specifically, the \code{Delta} class defines algorithms that \code{restrict} the values from a pair of
overlapping subdomains to their intersection.  This fundamental operation borrowed from the {\em sheaf} theory
(see, e.g., \cite{Bredon97})
allows us to detect \code{Map} cochains that agree on their overlaps.  Moreover (and this is a uniquely
computational feature), \code{Delta} allows us to \code{fuse} the values on the overlap back into the 
corresponding local \code{Maps} so as to ensure that they agree on the overlap and define a valid global map.
The \code{restrict}-\code{fuse} combination is a ubiquitous operation called \code{completion}, which we illustrate
here in detail in the case of {\em distributed} \code{Overlap} and \code{Delta}. For example, in
Section~\ref{sec:serial} we use \code{completion} to enforce the consistency of cones over points related by the overlap.

If the domain of the cochain \code{Map} carries no topology --- no connectivity between the points --- it
is simply a set and need not be represented by a \code{Sieve}.  This is the case for a pure linear algebra object, such
as a PETSc \code{Vec}. However, the \code{Overlap} and \code{Delta} still contain essential information about the
relationship among the subdomains and the data over them, and must be represented and constructed explicitly. In fact, a
significant part of an implementation of any domain decomposition problem should be the specification of the
\code{Overlap} and \code{Delta} pair, as they are at the heart of the problem.

Observe that \code{Overlap} fulfills \code{Sieve} functions at a larger scale, 
encoding the domain topology at the level of subdomains.  In fact, \code{Overlap} can be thought of as the ``superarrows''
between domain ``superpoints.''  Thus, the essential ideas of encoding topology by arrows indicating overlap between
pieces of the domain is the central idea behind the Sieve interface. Likewise, \code{Deltas} act as \code{Maps} on a
larger scale and can be \code{restricted} in accordance with an \code{Overlap}.

\subsection{Database interpretation}
The arrow-centric formalism of Sieve and the basic operations have an interpretations in terms of relational
databases and the associated `entity-relation' analyses.  Indeed, Sieve points can naturally be interpreted
as the rows of a table of `entities' (both in the database sense and the sense of `topological entity') with the point
itself serving as the key.  Arrows encode covering relations between points, and therefore define a natural binary
database relation with the composite key consisting of the two involved points. In this scenario cones and supports
have various interpretations in terms of queries against such a schema; in particular, the \code{cone} can be viewed as the
result of a (database) join of the arrow table with the point table on the target key; the \code{support} is the join
with the source key.  More interestingly, the topological \code{closure} is the transitive closure of the database
join applied to the arrow table; similarly for \code{star}.  Moreover, \code{meet} and \code{join} in the topological
sense cannot be formulated quite as succinctly in terms of database queries, but are very clear in terms of the
geometric intuitive picture of Sieve.

This can be contrasted with the scenario, in which only point entity tables are present and the covering or incident
points are stored in the entity record alongside the point key.  In this case, however, arrows have no independent
existence, are incapable of carrying their own ancillary information and are duplicated by each of the two related
points.  While in this paper we do not focus on the applications of arrow-specific data that can be attached to the
arrow records for lack of space, we illustrate its utility with a brief sketch of an example.

In extracting the cone or the (topological) closure of a point, such as a hexahedron in a 3D hex mesh, it is frequently
important to traverse the resulting faces, edges and points in the order determined by the orientation of the covered 
hex.  Each face, except those on the boundary, cover two hexahedra and most edges and vertices cover several faces and
edges, respectively.  Each of those covering relations induces a different orientation on the face, edge or vertex.
In FEM applications this results in a change of the sign of integral over the covering point. The sign, however, is not
intrinsically associated with the covering point, by rather with its orientation relative to the orientation induced
by the covered entity.  Thus, the sign of the integral is determined by the (covering,covered) pair, that is, by the
arrow.  In a entity-only schema, at worst there would be no natural place for the orientation data, and at best it would
make for an awkward design and potentially lead to storage duplication.  More sophisticated uses of arrow-specific data
include general transformation of the data attached to points upon its pullback onto the covered points (consider, for
example, the restriction/prolongation multigrid operators).

To summarize, Sieve can be viewed as an interface defining a relational database with a very particular schema and a
limit query set.  This query set, however, allows for some operations that may be difficult to describe succinctly in the 
database language (topological \code{meet} and \code{join})).  Furthermore, by defining a \emph{restricted} database
of topological entities and relations, as opposed to a flexible one, Sieve potentially allows for more effective optimizations
of the runtime and storage performance behind the same interface.  These issues will be discussed elsewhere.

\section{Mesh Distribution}\label{sec:partition}

Before our mesh is distributed, we must decide on a suitable partition, for which there are many excellent
packages (see, e.g.,~\cite{KarypisKumar98,parmetis-web-page,Chaco95}). We first construct suitable overlap
\code{Sieves}. The points will be abstract ``partitions'' that represent the sets of cells in each partition, with the
arrows connecting abstract partitions on different processes. The \code{Overlap} is used to structure the communication
of \code{Sieve} points among processes since the algorithm operates only on \code{Sections}, in this case we exhibit the
mesh \code{Sieve} as a \code{Section} with values in the space of points.

  \subsection{Dual Graph\label{subsec:dual-graph} and Partition encoding}

\begin{figure}
  \begin{center}
  \includegraphics[width=4.5in]{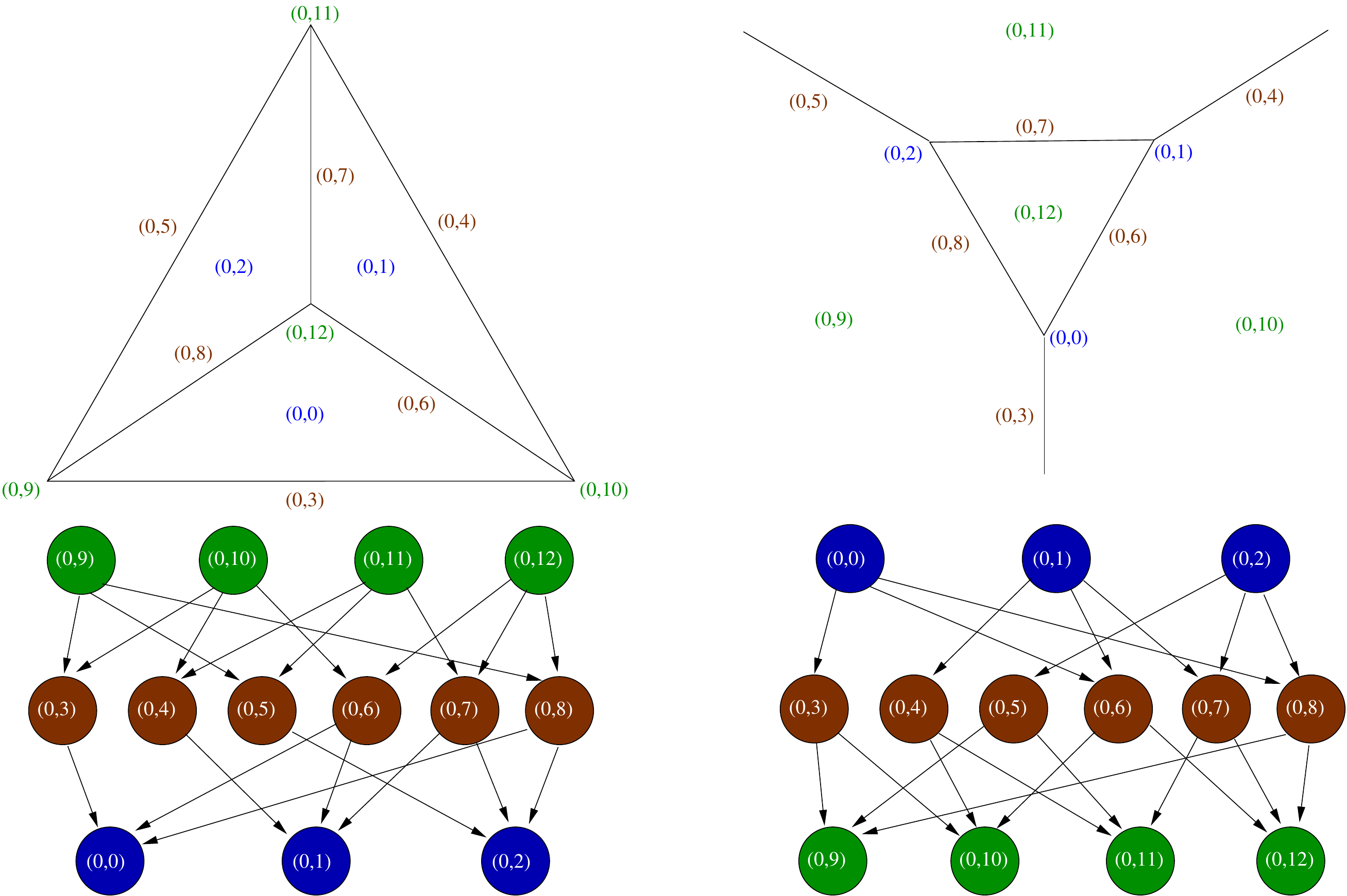}
  \end{center}
  \caption{A simple mesh and its dual.}
  \label{fig:dualSieve}
\end{figure}

    The graph partitioning algorithms in most packages, for example ParMetis and Chaco which were used for testing,
require the dual to our original mesh, sometimes referred to as the element connectivity graph. These packages partition
vertices of a graph, but FEM computations are best load-balanced by partitioning elements. Consider the simple mesh and
its dual, shown in Fig.~\ref{fig:dualSieve}. The dual Sieve is identical to the original except that all arrows are
reversed. Thus, we have an extremely simple characterization of the dual.

It is common practice to omit intermediate elements in the \code{Sieve}, for instance storing only cells and vertices. In this
case, we may construct the dual edges on the fly by looping over all cells in the mesh, taking the \code{support}, and placing
a dual edge for any support of the correct size (greater than or equal to the dimension is sufficient) between the two
cells in the support. Note this algorithm also works in parallel because the supports will, by definition, be identical
on all processes after support completion. Moreover, it is independent of the cell shape and dimension, unless the dual
edges must be constructed.

The partitioner returns an assignment of cells, vertices in the dual, to partitions. This can be thought of as a
\code{Section} over the mesh, giving the partition number for each cell. However, we will instead interpret this
assignment as a \code{Section} over the abstract partition points taking values in the space of \code{Sieve} points,
which can be used directly in our generic \code{Section} completion routine, described in
Section~\ref{subsec:completion}. In fact, Sieve can generate a partition of mesh elements of any dimension, for example
mesh faces in a finite volume code, using a hypergraph partitioner, such as that found in Zoltan~\cite{Zoltan06} and exactly the same
distribution algorithm.

\subsection{Distributing a Serial Mesh}\label{sec:serial}

    To make sense of a finite element mesh, we must first introduce a few new classes. A \code{Topology} combines a
sequence of \code{Sieves} with an \code{Overlap}. Our \code{Mesh} is modeled on the fiber bundle abstraction from
topology. Analogous to a topology combined with a fiber space, a \code{Mesh} combines a \code{Topology} with a
sequence of \code{Sections} over this topology. Thus, we may think of a \code{Mesh} as a \code{Topology} with several
distinguished \code{Sections}, the most obvious being the vertex coordinates.

    After the topology has been partitioned, we may distribute the \code{Mesh} in accordance with it, following the
steps below:
\begin{enumerate}
  \item Distribute the \code{Topology}.

  \item Distribute maps associated to the topology.

  \item Distribute bundle sections.
\end{enumerate}

Each distribution is accomplished by forming a specific \code{Section}, and then distributing that \code{Section} in
accordance with a given overlap. We call this process \emph{section completion}, and it is responsible for all
communication in the Sieve framework. Thus, we reduce parallel programming for the Sieve to defining the correct \code{Section}
and \code{Overlap}, which we discuss below.

  \subsubsection{Section Completion\label{subsec:completion}}

    Section completion is the process of completing cones, or supports, over a given overlap. Completion means
that the \code{cone} over a given point in the \code{Overlap} is sent to the \code{Sieve} containing the neighboring point, and
then fused into the existing cone of that neighboring point. By default, this fusion process is just insertion, but any binary
operation is allowed. For maximum flexibility, this operation is not carried out on global \code{Sections}, but rather on the
restriction of a \code{Section} to the \code{Overlap}, which we term \emph{overlap sections}. These can then be used to update the
global \code{Section}.

    The algorithm uses a recursive approach based on our decomposition of a \code{Section} into an atlas and
data. First the atlas, also a \code{Section}, is distributed, allowing receive data sizes to be calculated. Then the data
itself is sent. In this algorithm, we refer to the atlas, and its equivalent for section adapters, as a
\emph{sizer}. Here are the steps in the algorithm:
\begin{enumerate}
  \item Create send and receive sizer overlap sections.

  \item Fill send sizer section.

  \item Communicate.

  \item Create send and receive overlap sections.

  \item Fill send section.

  \item Communicate.
\end{enumerate}
The recursion ends when we arrive at a \code{ConstantSection}, described in~\cite{KnepleyKarpeev07C}, which does not have to be
distributed because it has the same value on every point of the domain.

  \subsubsection{Sieve Construction\label{subsec:construction}}

    The distribution process uses only section completion to accomplish all communication and data movement. We use
adapters~\cite{GammaHelmJohnsonVlissides95} to provide a \code{Section} interface to data, such as the partition. The
\code{PartitionSizeSection} adapter can be restricted to an abstract partition point, returning the total number of
sieve points in the partition (not just the those divided by the partitioner). Likewise, the \code{PartitionSection}
returns the points in a partition when restricted to the partition point. When we complete this section, the points are
distributed to the correct processes. All that remains is to establish the correct hierarchy among these points, which
we do by establishing the correct cone for each point. The \code{ConeSizeSection} and \code{ConeSection} adapters for
the \code{Sieve} return the cone size and points respectively when restricted to a point. We see here that a sieve
itself can be considered a section taking values in the space of points. Thus sieve completion consists of the following:
\begin{enumerate}
  \item\label{step:localCopy} Construct local mesh from partition assignment by copying.

  \item\label{step:initOverlap} Construct initial partition overlap.

  \item Complete the partition section to distribute the cells.

  \item\label{step:sieveOverlap} Update the \code{Overlap} with the points from the overlap sections.

  \item\label{step:coneCompletion} Complete the cone section to distribute remaining \code{Sieve} points.

  \item Update local \code{Sieves} with \code{cones} from the overlap sections.
\end{enumerate}
The final \code{Overlap} now relates the parallel \code{Sieve} to the initial serial \code{Sieve}. Note that we have
used only the \code{cone()} primitive, and thus this algorithm applies equally well to meshes of any dimension, element
shape, or connectivity. In fact, we could distribute an arbitrary graph without changing the algorithm.

\subsection{Redistributing a Mesh}\label{sec:redistribution}

    Redistributing an existing parallel mesh is identical to distributing a serial mesh in our framework. However, now
the send and receive \code{Overlaps} are potentially nonempty for every process. The construction of the intermediate
partition and cone \code{Sections}, as well as the section \code{completion} algorithm, remain exactly as before. Thus,
our high level of abstraction has resulted in enormous savings through code reuse and reduction in complexity.

\begin{figure}
  \begin{center}
  \includegraphics[height=5.0in]{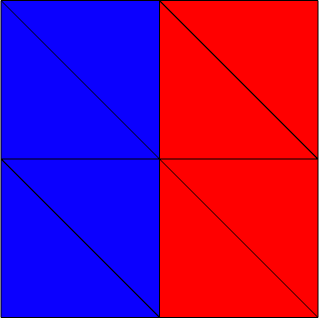}
  \end{center}
  \caption{Initial distributed triangular mesh.}
  \label{fig:triMeshDistBefore}
\end{figure}

\begin{figure}
  \begin{center}
  \includegraphics[width=2.5in]{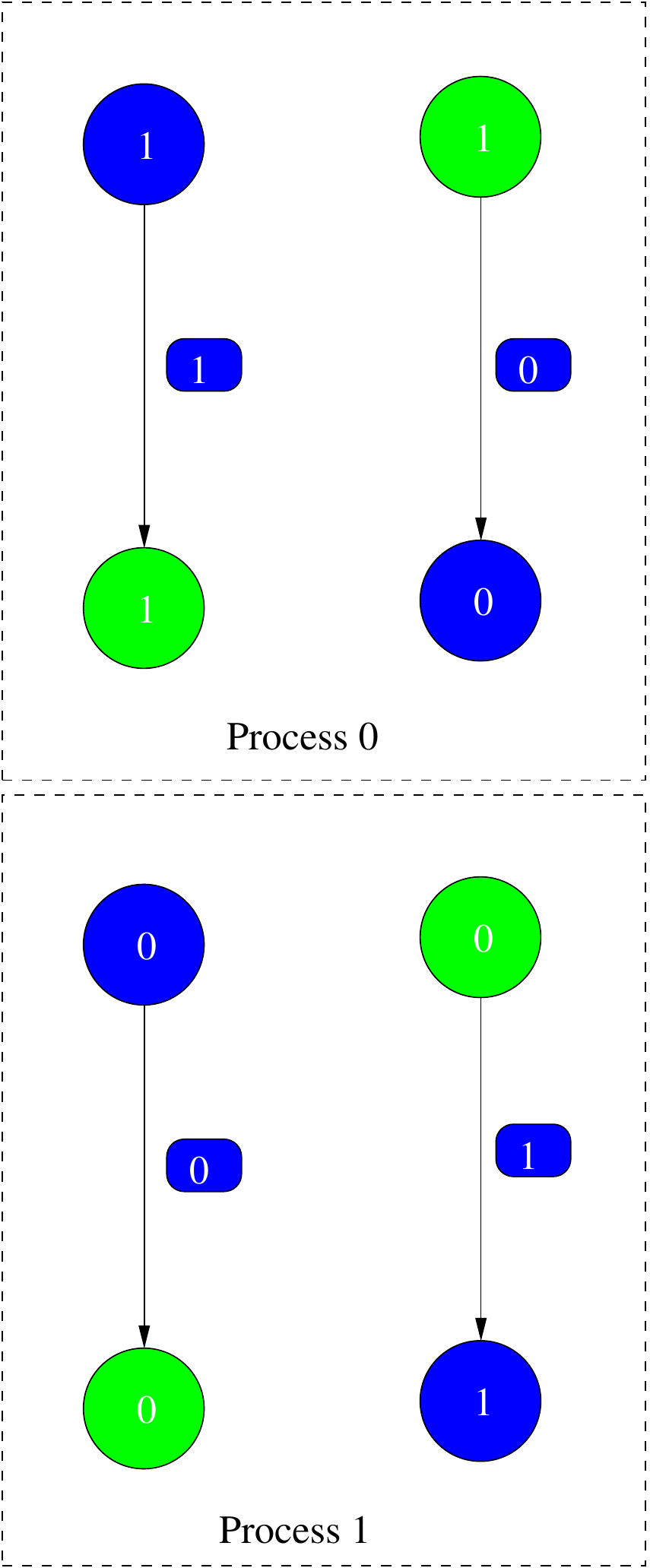}
  \end{center}
  \caption{Partition point \code{Overlap}, with dark partition points, light process ranks, and \code{arrow} labels
representing remote points. The send \code{Overlap} is on the left, and the receive \code{Overlap} on the right.}
  \label{fig:partOverlapRedist}
\end{figure}

\begin{figure}
  \begin{center}
  \includegraphics[width=3.0in]{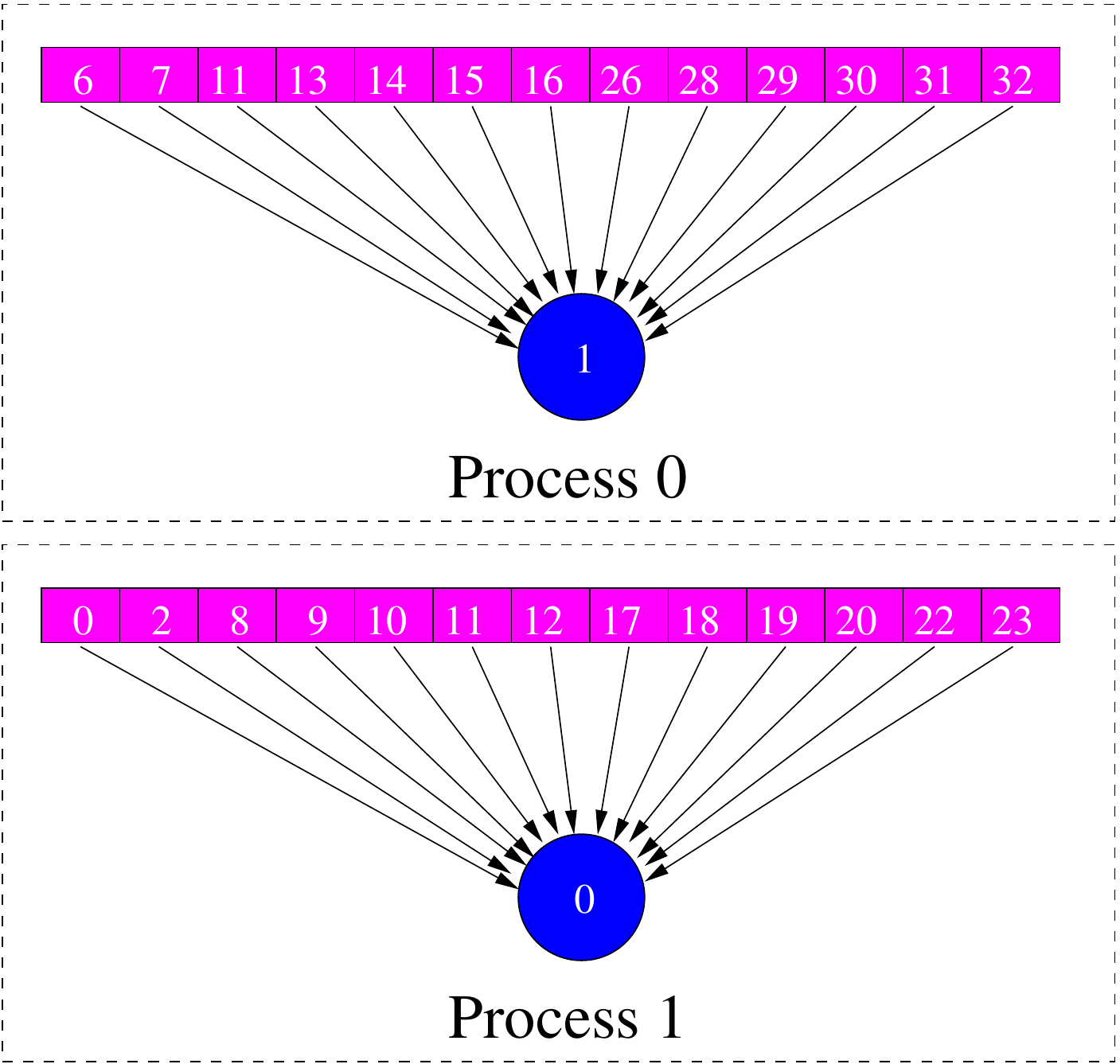}
  \end{center}
  \caption{Partition section, with circular partition points and rectangular Sieve point data.}
  \label{fig:partSectionRedist}
\end{figure}

\begin{figure}
  \begin{center}
  \includegraphics[height=7.3in]{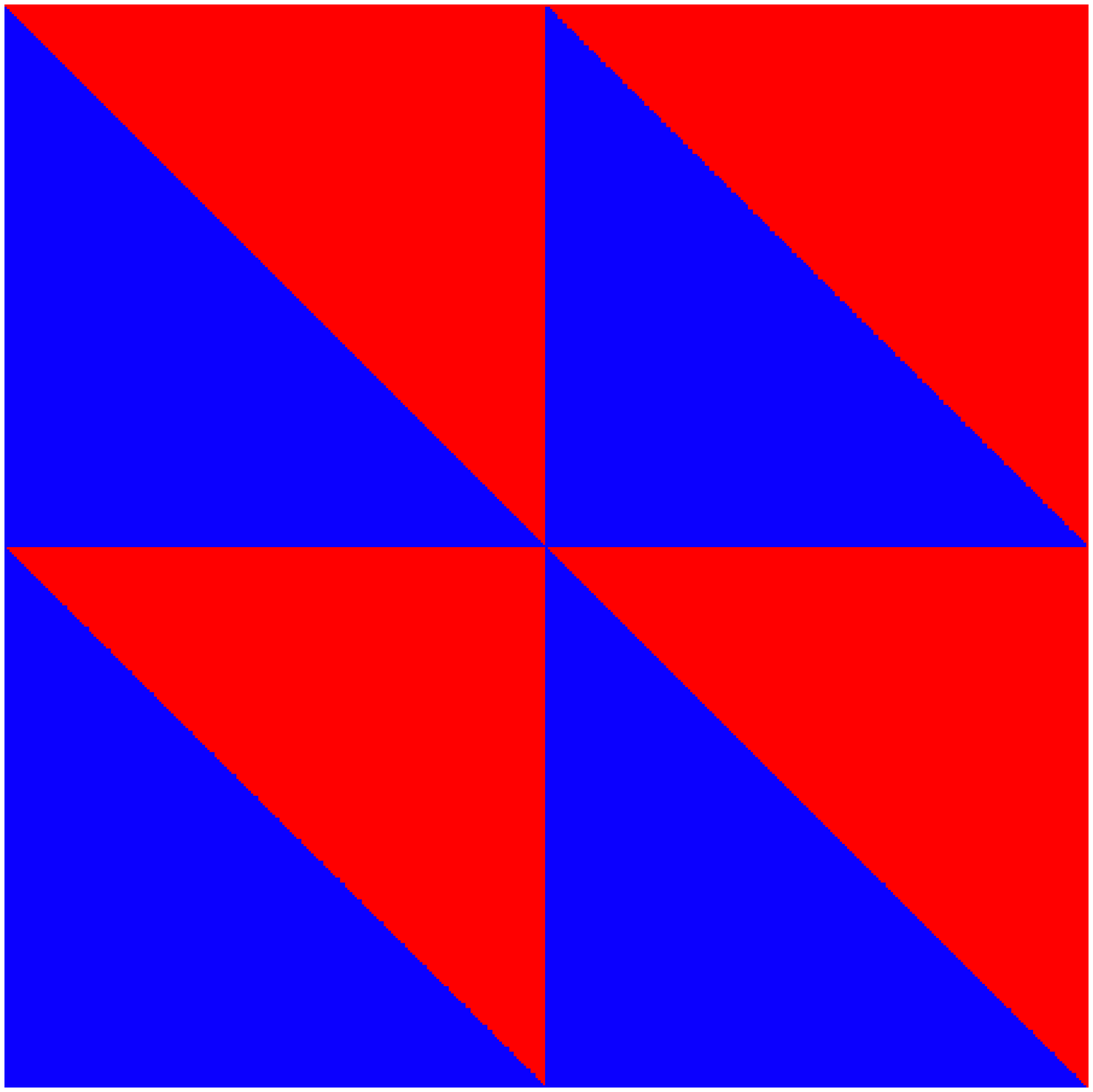}
  \end{center}
  \caption{Redistributed triangular mesh.}
  \label{fig:triMeshDistAfter}
\end{figure}

    As an example, we return to the triangular mesh discussed earlier. However, we will begin with the distributed mesh shown in
Fig.~\ref{fig:triMeshDistBefore}, which assigns triangles (4, 5, 6, 7) to process 0, and (0, 1, 2, 3) to process 1. 
The main difference in this example will be the \code{Overlap}, which determines the communication pattern. In
Fig.~\ref{fig:partOverlapRedist}, we see that each process will both send and receive data during the redistribution.
Thus, the partition \code{Section} in Fig.~\ref{fig:partSectionRedist} has data on both processes. Likewise, upon
\code{completion} we can construct a Sieve \code{Overlap} with both send and receive portion on each process. Cone and
coordinate \code{completion} also proceed exactly as before, except that data will flow between both processes. We
arrive in the end at the redistributed mesh shown in Fig.~\ref{fig:triMeshDistAfter}. No operation other than Section
\code{completion} itself was necessary.

\section{Examples\label{subsec:distributionEx}}

\begin{figure}
  \begin{center}
  \includegraphics[width=4.5in]{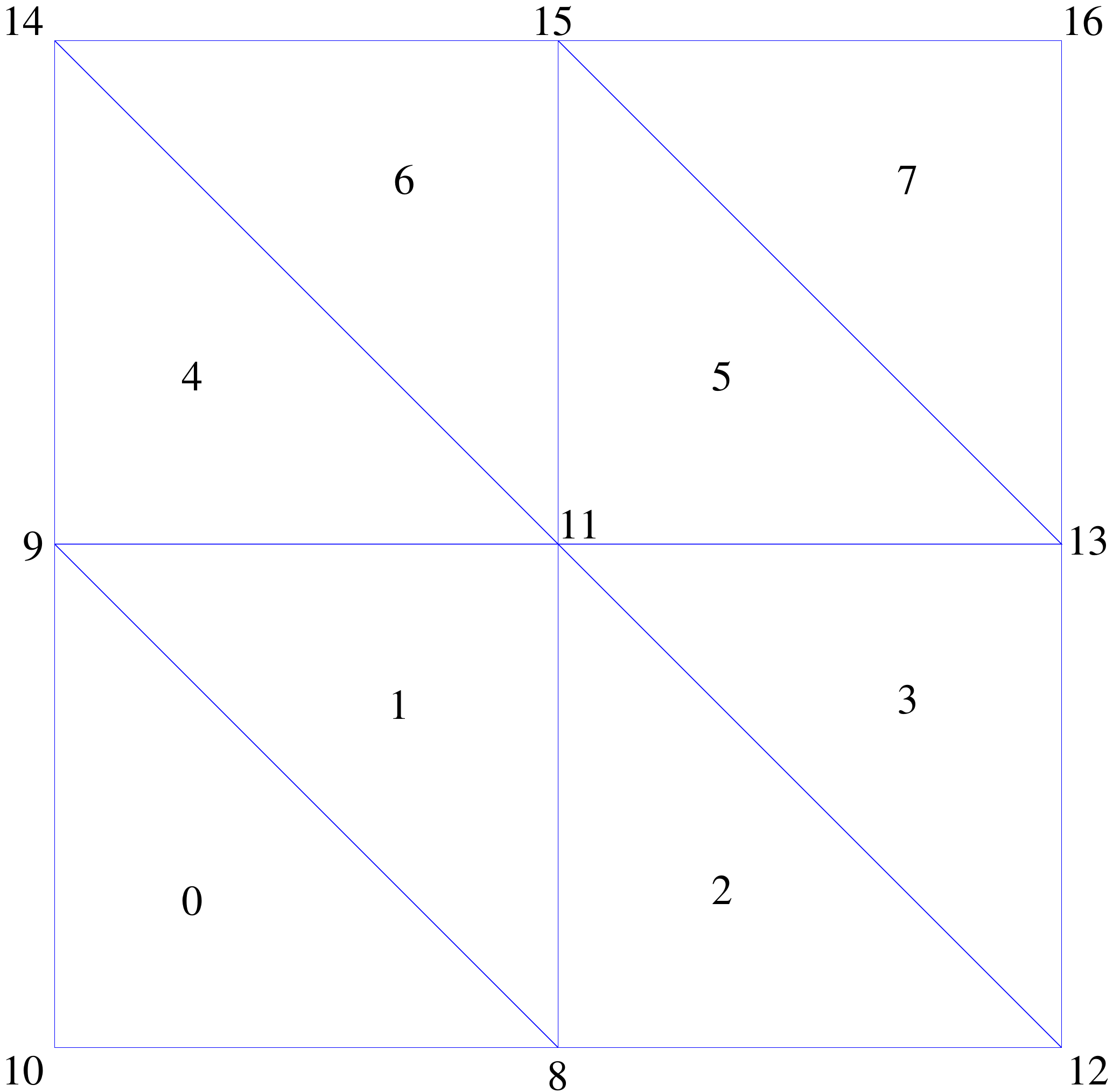}
  \end{center}
  \caption{A simple triangular mesh.}
  \label{fig:triangularMesh}
\end{figure}

\begin{figure}
  \begin{center}
  \includegraphics[width=4.5in]{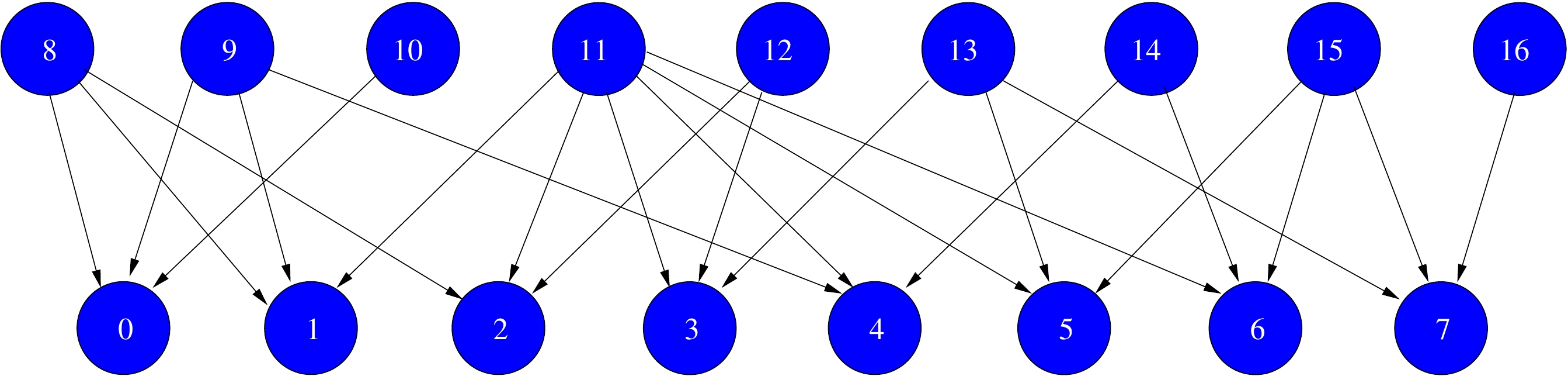}
  \end{center}
  \caption{\code{Sieve} for mesh in Fig.~\ref{fig:triangularMesh}.}
  \label{fig:triangularSieve}
\end{figure}

\begin{figure}
  \begin{center}
  \includegraphics[width=4.5in]{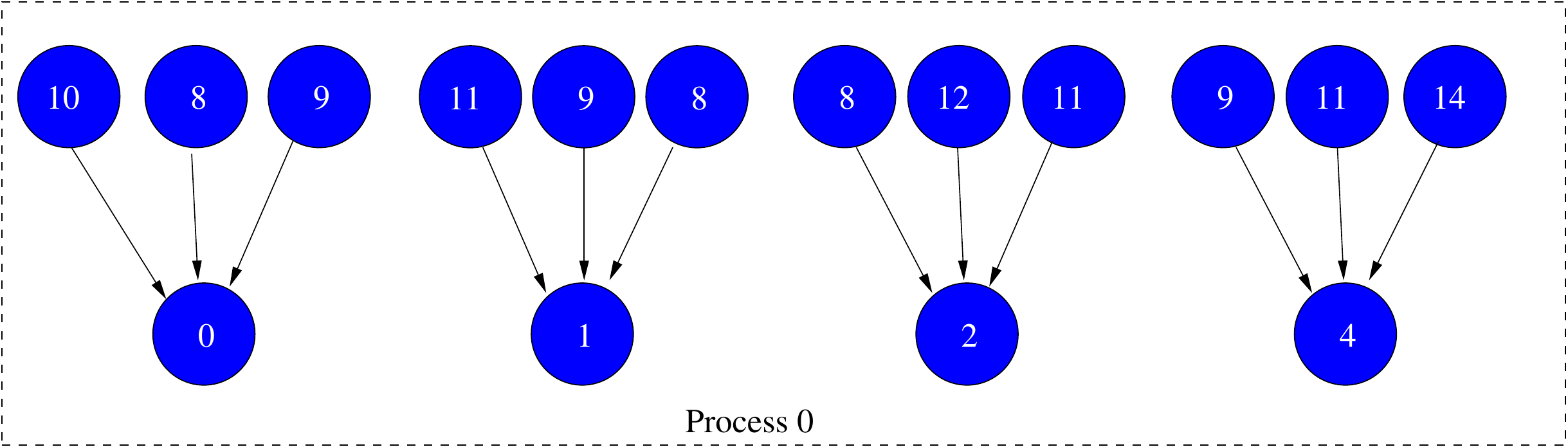}
  \end{center}
  \caption{Initial local sieve on process 0 for mesh in Fig.~\ref{fig:triangularMesh}.}
  \label{fig:triangularSieve_local}
\end{figure}

    To illustrate the distribution method, we begin with a simple square triangular mesh, shown in
Fig.~\ref{fig:triangularMesh} with its corresponding \code{Sieve} shown in Fig.~\ref{fig:triangularSieve}. We
distribute this mesh onto two processes: the partitioner assigns triangles (0, 1, 2, 4) to process 0, and (3, 5, 6, 7)
to process 1. In step~\ref{step:localCopy}, we create a local \code{Sieve} on process 0, shown in
Fig.~\ref{fig:triangularSieve_local}, since we began with a serial mesh.

\begin{figure}
  \begin{center}
  \includegraphics[width=2.5in]{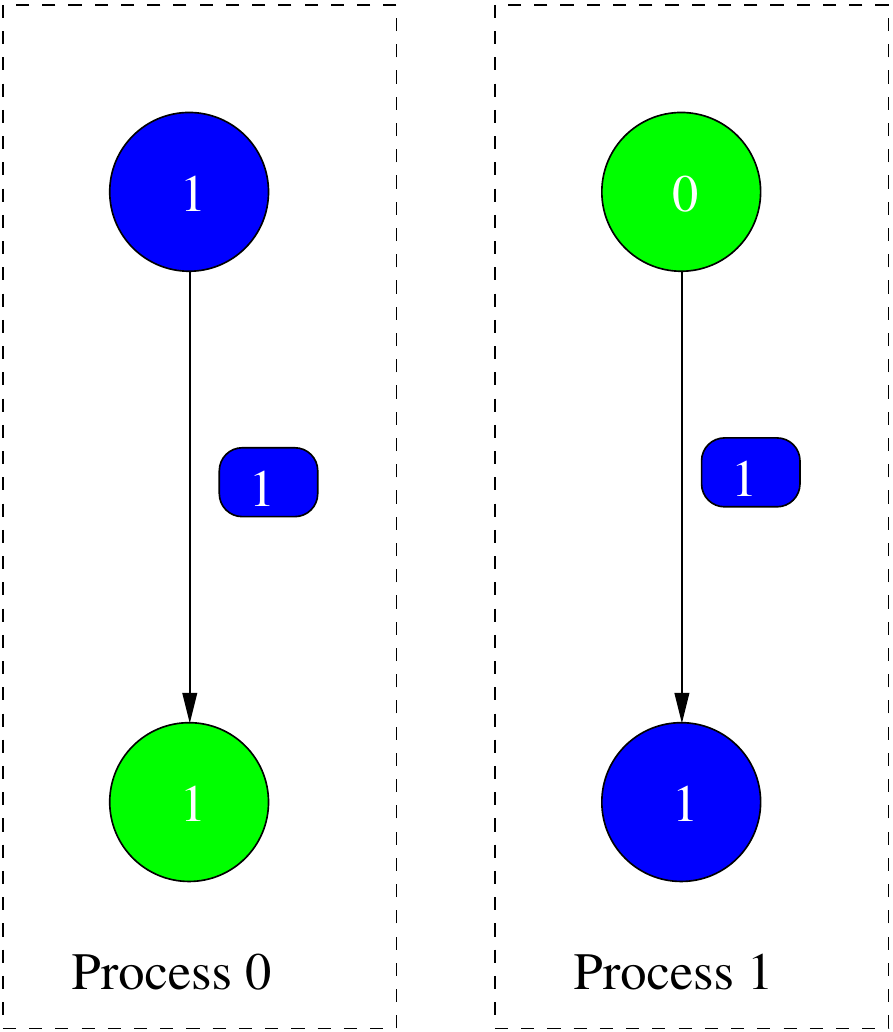}
  \end{center}
  \caption{Partition point \code{Overlap}, with dark partition points, light process ranks, and \code{arrow} labels
representing remote points.}
  \label{fig:partOverlap}
\end{figure}

    For step~\ref{step:initOverlap}, we identify abstract partition points on the two processes using an overlap
\code{Sieve}, shown in Fig.~\ref{fig:partOverlap}. Since this step is crucial to an understanding of the algorithm, we
will explain it in detail. Each \code{Overlap} is a \code{Sieve}, with dark circles representing abstract partition
points, and light circles process ranks. The rectangles are Sieve \code{arrow} data, or labels, representing remote
partition points. The send \code{Overlap} is shown for process 0, identifying the partition point 1 with the same point
on process 1. The corresponding receive \code{Overlap} is shown for process 1. The send \code{Overlap} for process 1 and
receive \code{Overlap} for process 0 are both null because we are broadcasting a serial mesh from process 0.

\begin{figure}
  \begin{center}
  \includegraphics[width=3.0in]{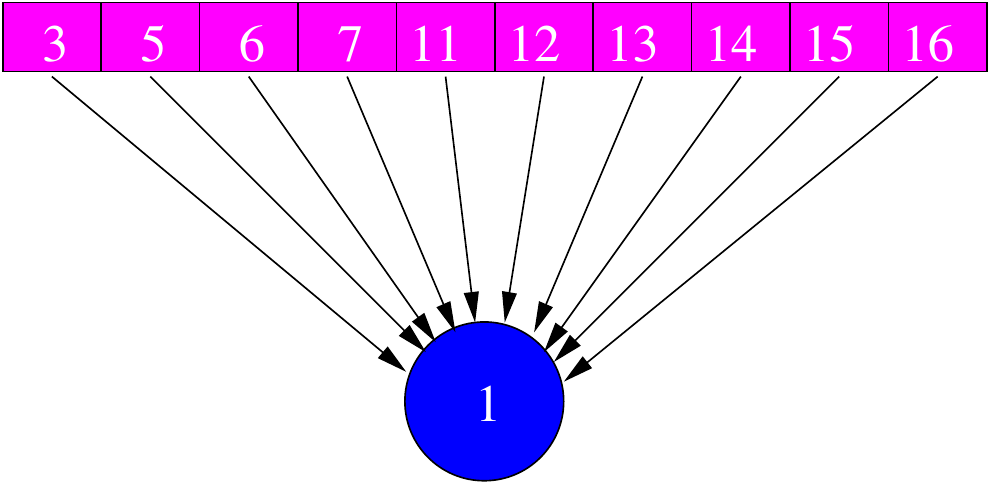}
  \end{center}
  \caption{Partition section, with circular partition points and rectangular Sieve point data.}
  \label{fig:partSection}
\end{figure}

    We now complete the partition \code{Section}, using the partition \code{Overlap}, in order to distribute the Sieve
points. This \code{Section} is shown in Fig.~\ref{fig:partSection}. Not only are the four triangles in partition 1
shown, but also the six vertices. The receive overlap \code{Section} has a base consisting of the overlap points, in
this case partition point 1; the cap will be completed, meaning that it now has the Sieve points in the cap.

\begin{figure}
  \begin{center}
  \includegraphics[width=4.5in]{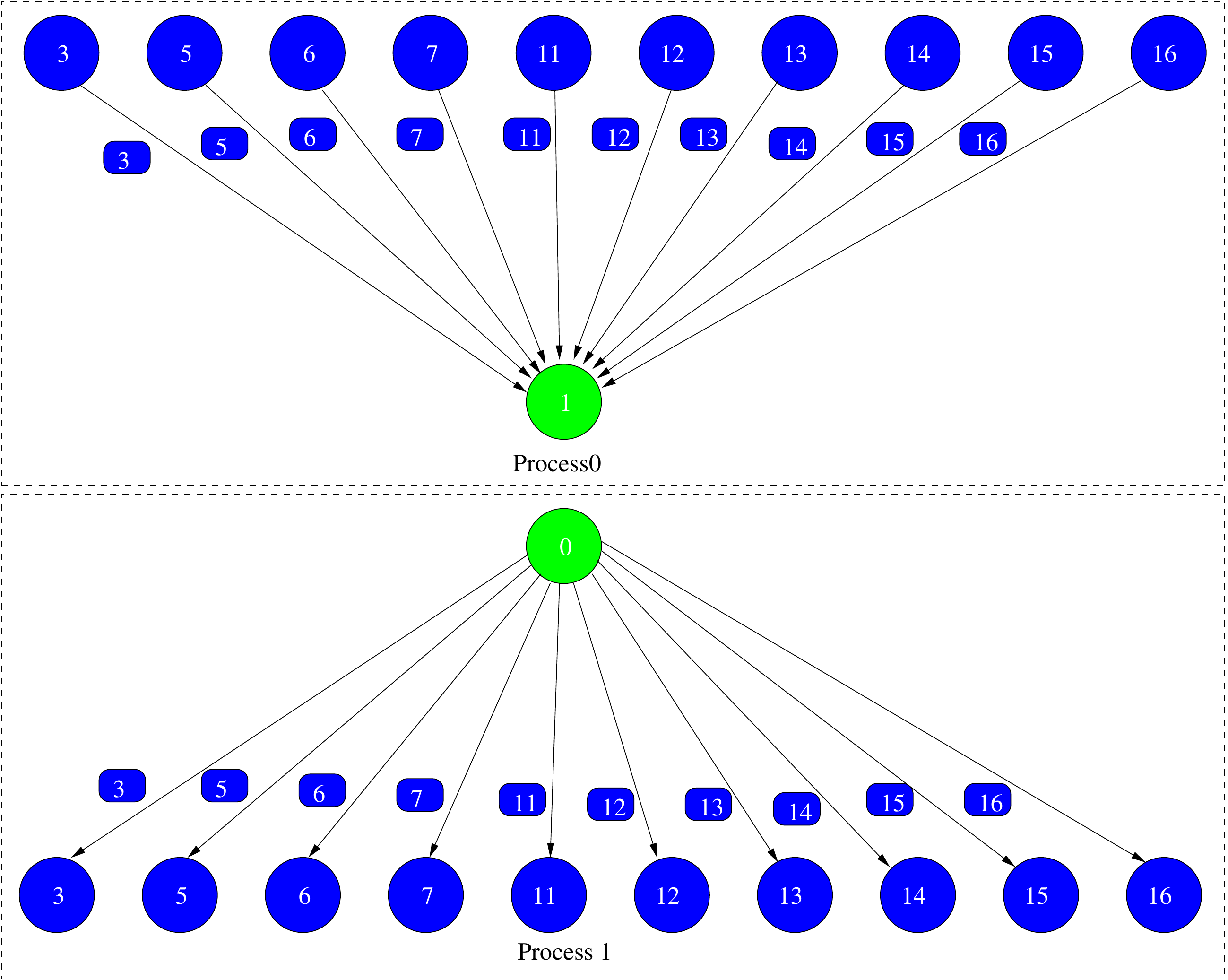}
  \end{center}
  \caption{Sieve overlap, with Sieve points in blue, process ranks in green, and \code{arrow} labels
representing remote sieve points.}
  \label{fig:sieveOverlap}
\end{figure}

\begin{figure}
  \begin{center}
  \includegraphics[width=3.0in]{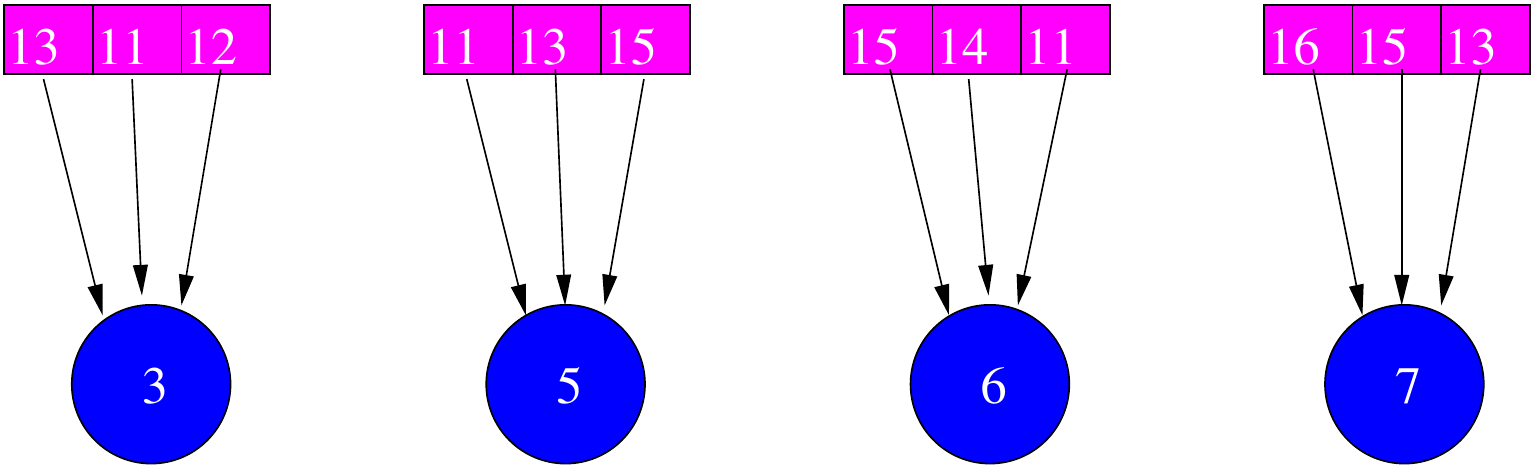}
  \end{center}
  \caption{Cone \code{Section}, with circular Sieve points and rectangular \code{cone} point data.}
  \label{fig:coneSection}
\end{figure}

\begin{figure}
  \begin{center}
  \includegraphics[width=4.5in]{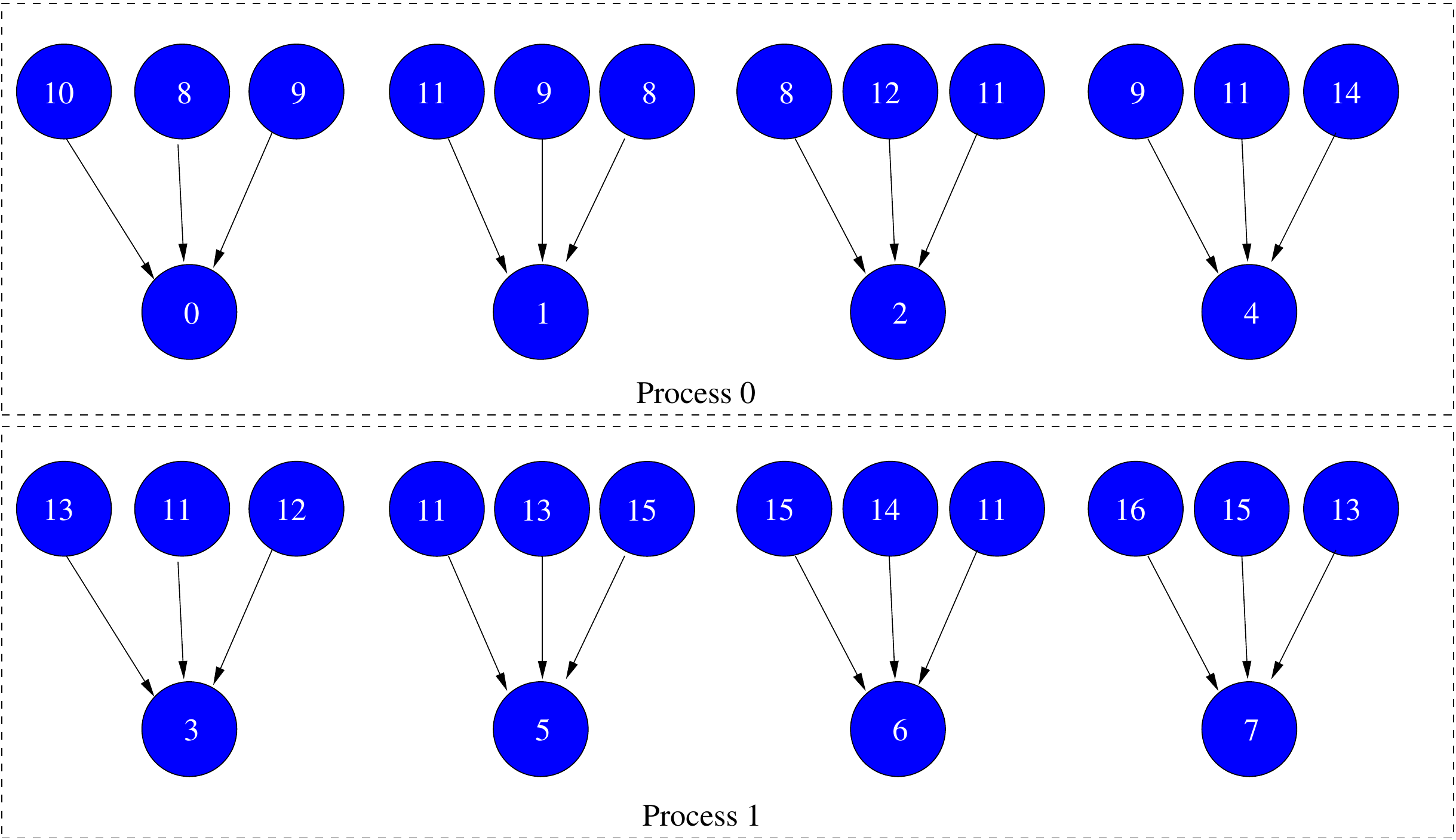}
  \end{center}
  \caption{Distributed Sieve for mesh in Fig.~\ref{fig:triangularMeshDist}.}
  \label{fig:triangularSieveDist}
\end{figure}

    Using the receive overlap \code{Section} in step~\ref{step:sieveOverlap}, we can update our \code{Overlap} with the
new Sieve points just distributed to obtain the \code{Overlap} for Sieve points rather than partition points. The Sieve
\code{Overlap} is shown in Fig.~\ref{fig:sieveOverlap}. Here identified points are the same on both processes, but this
need not be the case.
In step~\ref{step:coneCompletion} we complete the cone \code{Section}, shown in Fig.~\ref{fig:coneSection}, distributing
the covering relation. We use the \code{cones} in the receive overlap \code{Section} to construct the distributed Sieve
in Fig.~\ref{fig:triangularSieveDist}.

\begin{figure}
  \begin{center}
  \includegraphics[width=4.5in]{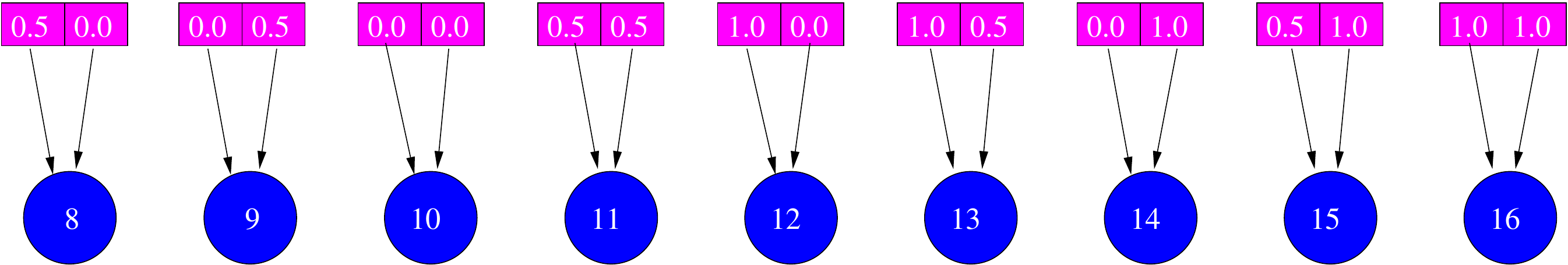}
  \end{center}
  \caption{Coordinate \code{Section}, with circular Sieve points and rectangular coordinate data.}
  \label{fig:coordSection}
\end{figure}

\begin{figure}
  \begin{center}
  \includegraphics[width=4.5in]{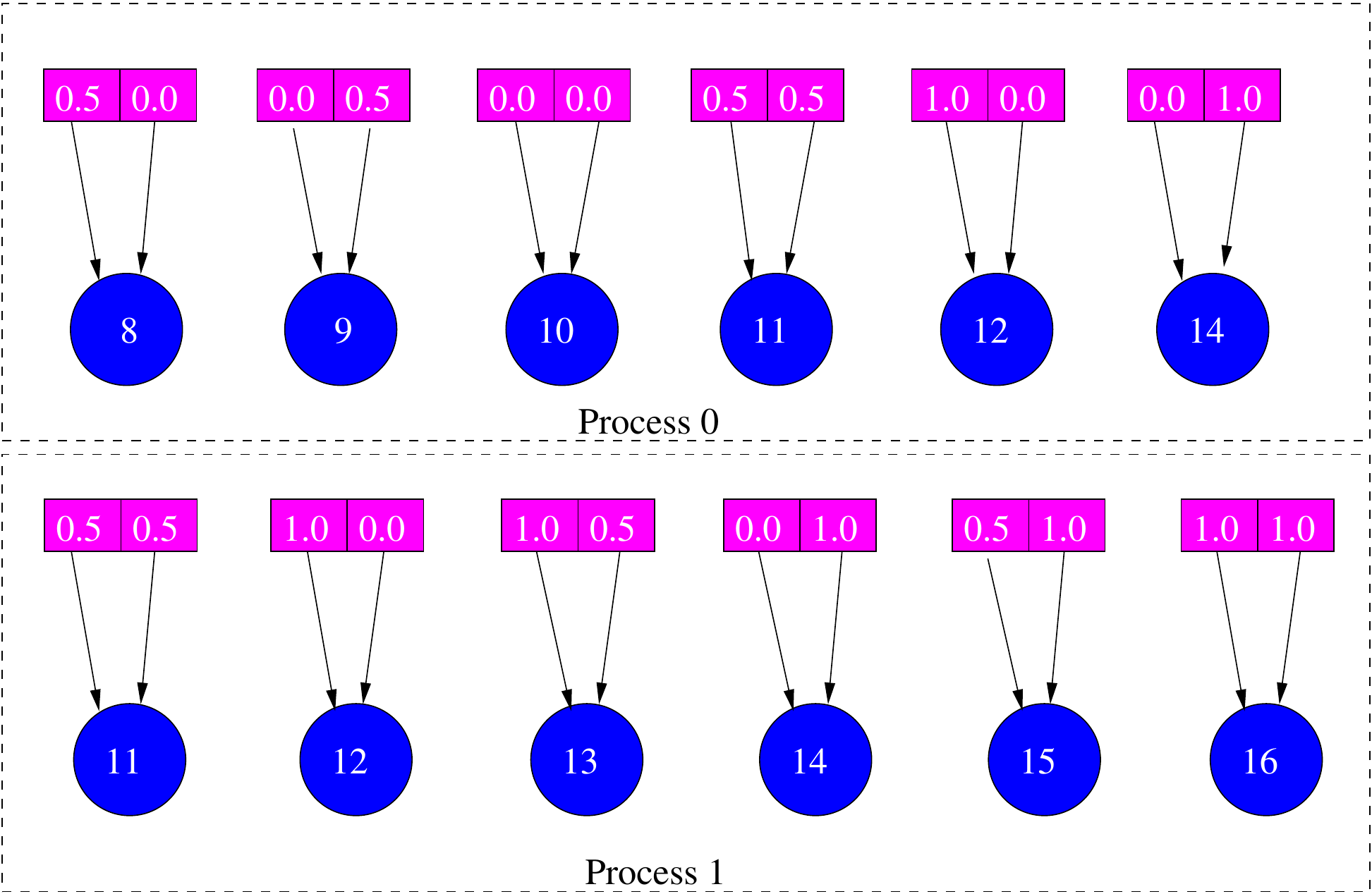}
  \end{center}
  \caption{Distributed coordinate \code{Section}.}
  \label{fig:coordSectionDist}
\end{figure}

\begin{figure}
  \begin{center}
  \includegraphics[width=4.5in]{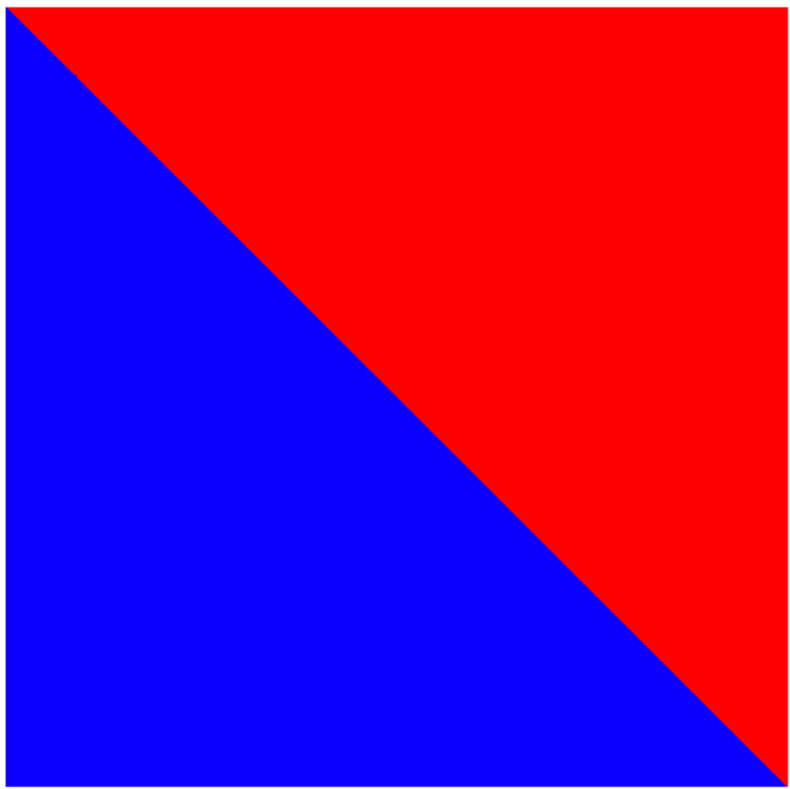}
  \end{center}
  \caption{The distributed triangular mesh.}
  \label{fig:triangularMeshDist}
\end{figure}

    After distributing the topology, we distribute any associated \code{Sections} for the \code{Mesh}. In this
example, we have only a coordinate \code{Section}, shown in Fig.~\ref{fig:coordSection}. Notice that while only vertices
have coordinate values, the Sieve \code{Overlap} contains the triangular faces as well. Our algorithm is insensitive to
this, as the faces merely have empty \code{cones} in this \code{Section}. We now make use of another adapter, the
\code{Atlas}, which substitutes the number of values for the values returned by a \code{restrict}, which we use as
the sizer for \code{completion}. After distribution of this \code{Section}, we have the result in
Fig.~\ref{fig:coordSectionDist}. We are thus able to fully construct the distributed mesh in
Fig.~\ref{fig:triangularMeshDist}.

\begin{figure}
  \begin{center}
  \includegraphics[width=4.5in]{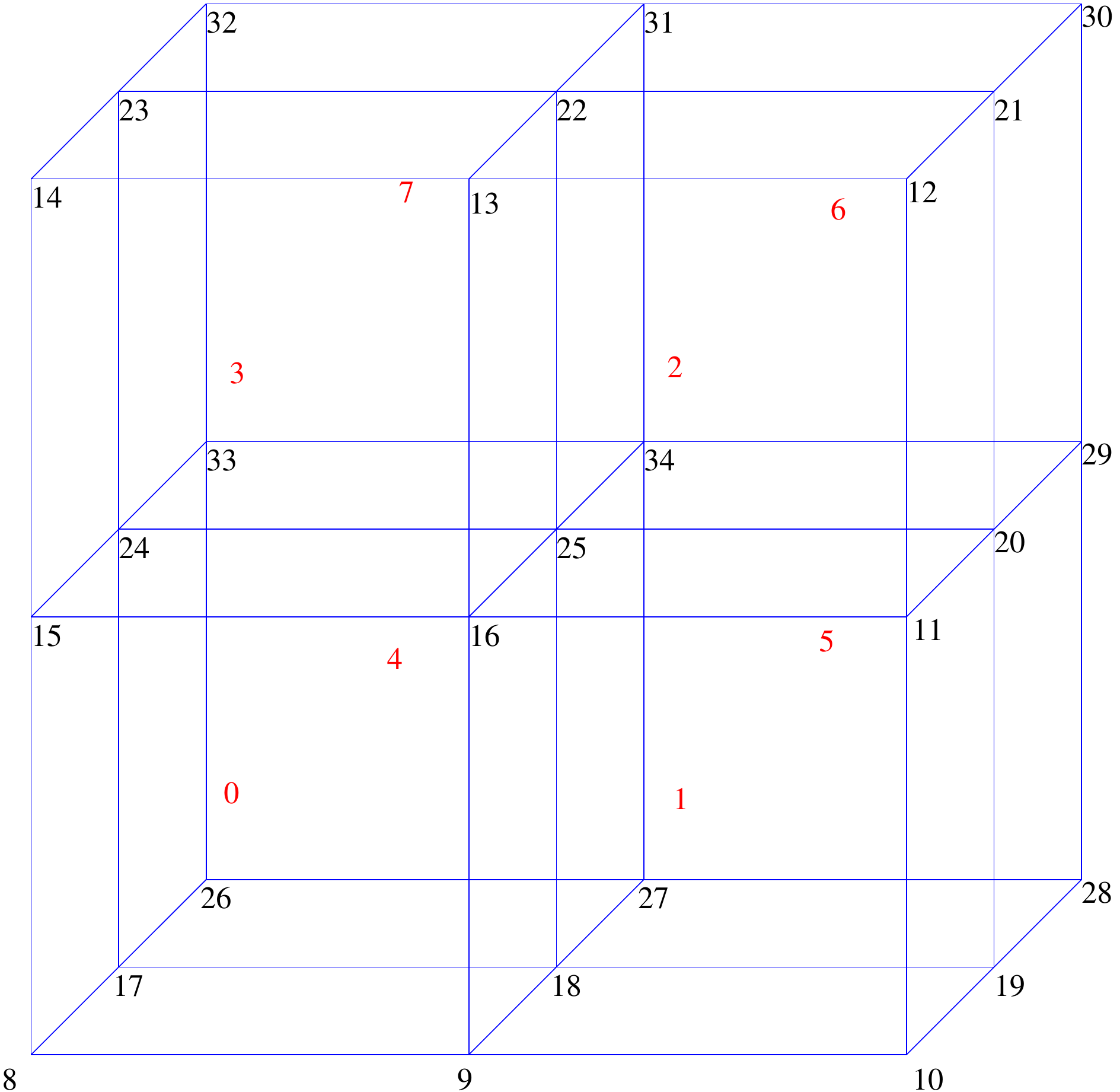}
  \end{center}
  \caption{A simple hexahedral mesh.}
  \label{fig:hexMesh}
\end{figure}

\begin{figure}
  \begin{center}
  \includegraphics[height=8.0in]{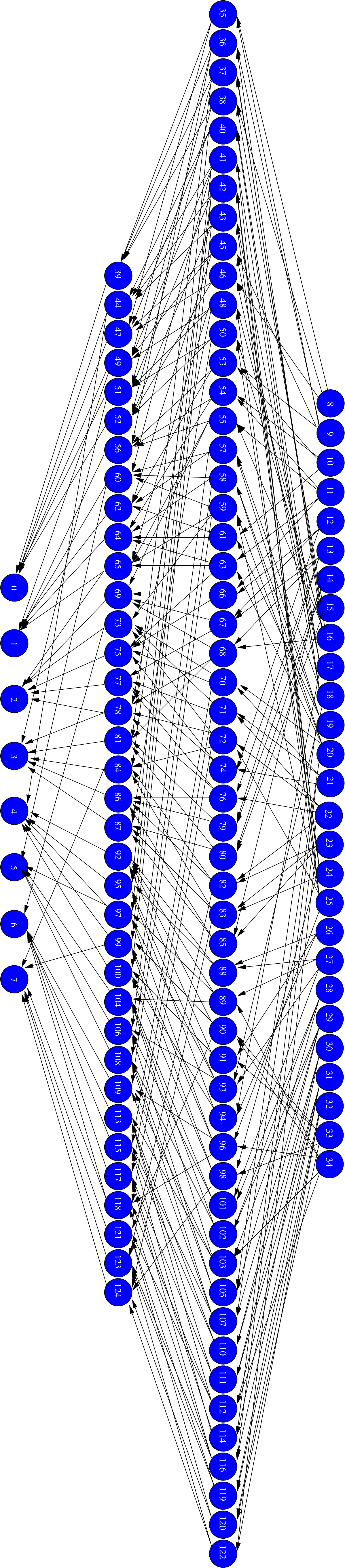}
  \end{center}
  \caption{Sieve corresponding to the mesh in Fig.~\ref{fig:hexMesh}.}
  \label{fig:hexSieve}
\end{figure}

    The mesh distribution method is independent of the topological dimension of the mesh, the embedding, the cell
shapes, and even the type of element determining the partition. Moreover, it does not depend on the existence of
intermediate mesh elements in the Sieve. We will change each of these in the next example, distributing a
three-dimensional hexahedral mesh, shown in Fig.~\ref{fig:hexMesh}, by partitioning the faces.
As one can see from Fig.~\ref{fig:hexSieve}, the Sieve is complicated even for this simple mesh. However, it does
have recognizable structures. Notice that it is stratified by the topological dimension of the points. This is a feature
of any cell complex when represented as a Sieve.

\begin{figure}
  \begin{center}
  \includegraphics[width=3.0in]{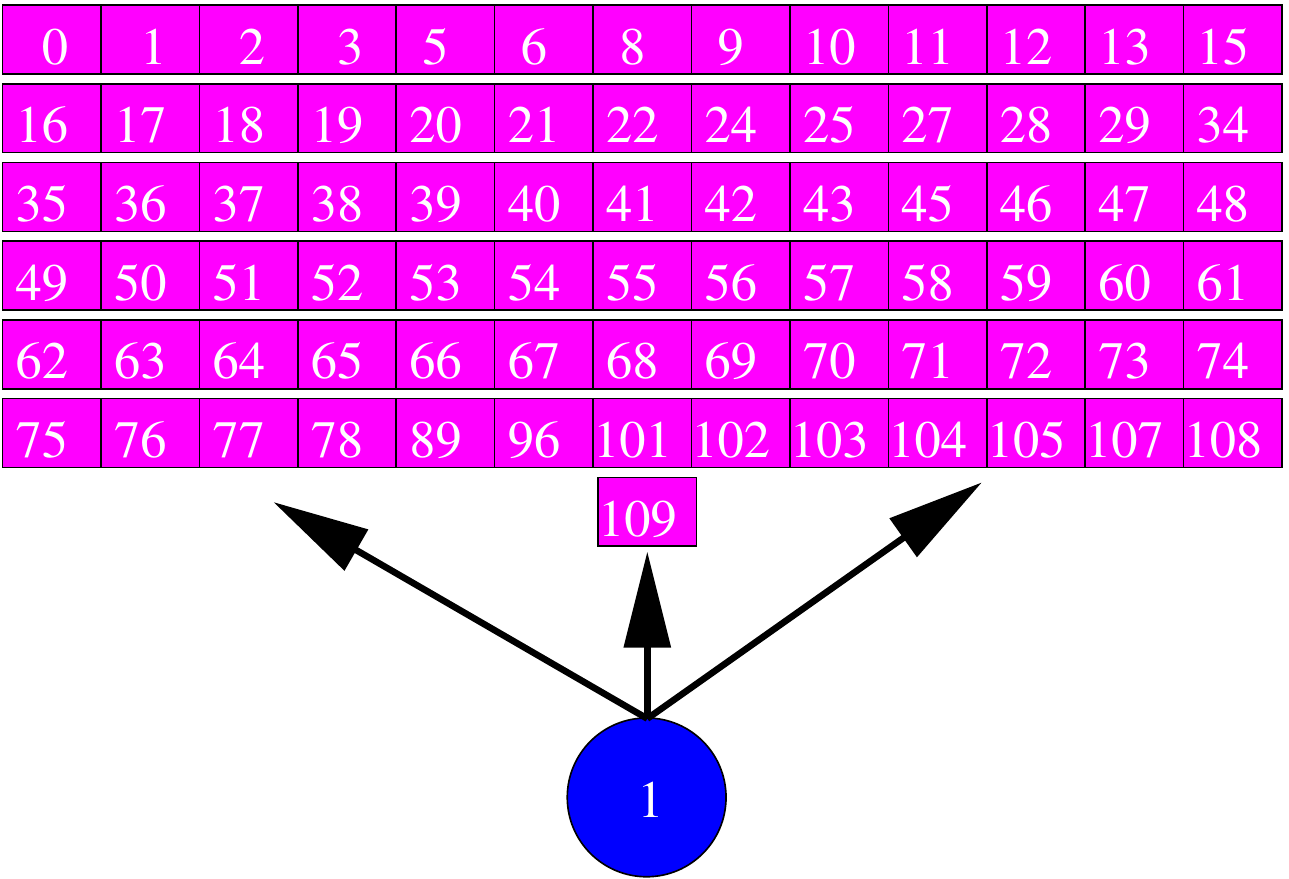}
  \end{center}
  \caption{Partition \code{Section}, with circular partition points and rectangular Sieve point data.}
  \label{fig:partSection3D}
\end{figure}

\begin{figure}
  \begin{center}
  \includegraphics[height=7.3in]{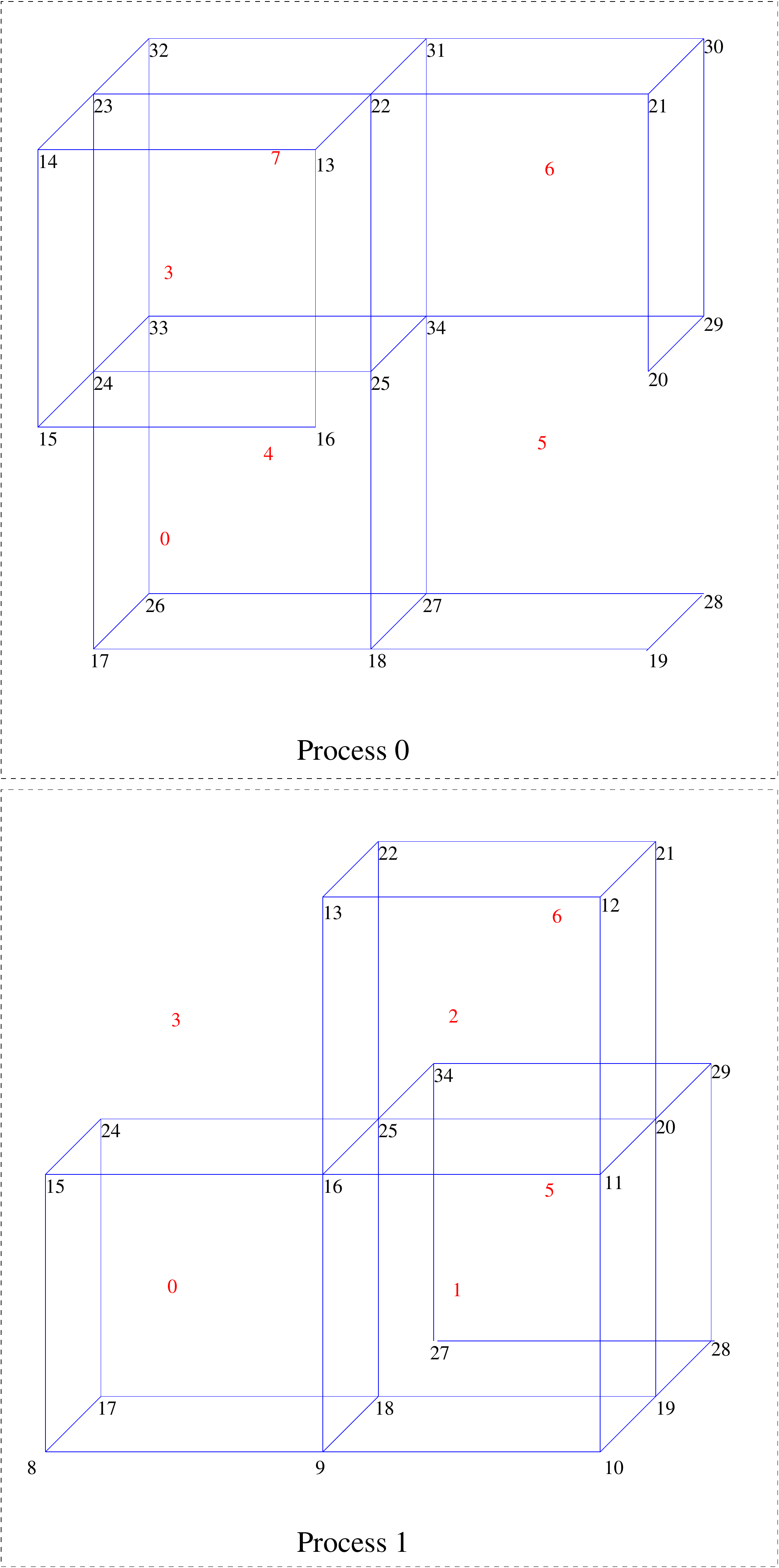}
  \end{center}
  \caption{Distributed hexahedral mesh.}
  \label{fig:hexMeshDist}
\end{figure}

    The partition \code{Overlap} in this case is exactly the one shown in Fig.~\ref{fig:partOverlap}; even though an
edge partition was used instead of the cell partition common for finite elements, the partition \code{Section} in
Fig.~\ref{fig:partSection3D} looks the same although with more data. Not only is the closure of the edges
included, but also their star. This is the abstract method to determine all points in a given partition. The Sieve
\code{Overlap} after \code{completion} is also much larger but has exactly the same structure.
In fact, all operations have exactly the same form because the section \code{completion} algorithm is independent of all
the extraneous details in the problem. The final partitioned mesh is shown in Fig.~\ref{fig:hexMeshDist}, where we see
that ghost cells appear automatically when we use a face partition.

\section{Conclusions}\label{sec:conclusions}

    We have presented mesh partitioning and distribution in the context of the Sieve framework in order to illustrate
the power and flexibility of this approach. Since we draw no distinction between mesh elements of any shape, dimension,
or geometry, we may accept a partition of any element type, such as cells or faces. Once provided with this partition
and an overlap sieve, which just indicates the flow of information and is constructed automatically, the entire mesh
can be distributed across processes by using a single operation, \emph{section completion}. Thus, only a single parallel
operation need be portable, verifiable, or optimized for a given architecture. Moreover, this same operation can be used
to distribute data associated with the mesh, in any arbitrary configuration, according to the same partition. Thus, the
high level of mathematical abstraction in the Sieve interface results in concrete benefits in terms of code reuse,
simplicity, and extensibility.

\section*{Acknowledgements}

The authors benefited from many useful discussions with Gary Miller and Rob Kirby. This work was supported by the
Mathematical, Information, and Computational Sciences Division subprogram of the Office of Advanced Scientific Computing
Research, Office of Science, U.S. Department of Energy, under Contract DE-AC02-06CH11357.

\bibliographystyle{plain}
\bibliography{petsc,petscapp}


\end{document}